\documentclass[lettersize,journal]{IEEEtran}
\usepackage{amsmath,amsfonts}
\usepackage{algorithmic}
\usepackage{algorithm}
\usepackage{array}
\usepackage[caption=false,font=normalsize,labelfont=sf,textfont=sf]{subfig}
\usepackage{textcomp}
\usepackage{stfloats}
\usepackage{url}
\usepackage{verbatim}
\usepackage{graphicx}
\usepackage{cite}
\usepackage{tikz}
\usepackage{pgfplots}
\usepackage[T1]{fontenc}

\usepackage{multirow}

\pgfplotsset{every axis/.append style={
                    label style={font=\small},
                    tick label style={font=\small},
                    title style={font=\small}
                    },
                    legend style={font=\small}
}

\newenvironment{customlegend}[1][]{%
    \begingroup
    \csname pgfplots@init@cleared@structures\endcsname
    \pgfplotsset{#1}%
}{%
    \csname pgfplots@createlegend\endcsname
    \endgroup
}%

\def\addlegendimage{\csname pgfplots@addlegendimage\endcsname}

\usetikzlibrary{calc,positioning,shapes.geometric}

 \pgfplotsset{compat=1.3} 

\newcommand{\filepath}[1]{{\tt #1}}
\usetikzlibrary{narrow}

\graphicspath{ {./images/} }
\definecolor{mycolor4}{RGB}{230,97,1}
\definecolor{mycolor2}{RGB}{178,171,210}
\definecolor{mycolor3}{RGB}{253,184,99}
\definecolor{mycolor1}{RGB}{94,60,153}

\begin{document}

\title{GreenEVT: Greensboro Electric Vehicle Testbed}

\author{Gustav Nilsson,~\IEEEmembership{Member,~IEEE,} Alejandro D. Owen Aquino,~\IEEEmembership{Student Member,~IEEE,}\\ Samuel Coogan,~\IEEEmembership{Senior Member,~IEEE,} and Daniel K. Molzahn~\IEEEmembership{Senior Member,~IEEE}
\thanks{
 G. Nilsson is with the School of Architecture, Civil and Environmental Engineering, École Polytechnique Fédérale de Lausanne (EPFL), 1015 Lausanne, Switzerland. {gustav.nilsson@epfl.ch}.
 A.D.~Owen Aquino, D.K.~Molzahn, and S.~Coogan are with the School of Electrical and Computer Engineering, Georgia Institute of Technology, Atlanta, 30332, USA. \{aaquino, molzahn, sam.coogan\}@gatech.edu. S.~Coogan is also with the School of Civil and Environmental Engineering, Georgia Institute of Technology. This work was supported by the Strategic Energy Institute at Georgia Tech.}%
}



\maketitle

\begin{abstract}
The ongoing electrification of the transportation fleet will increase the load on the electric power grid. Since both the transportation network and the power grid already experience periods of significant stress, joint analyses of both infrastructures will most likely be necessary to ensure acceptable operation in the future. To enable such analyses, this paper presents an open-source testbed that jointly simulates high-fidelity models of both the electric distribution system and the transportation network. The testbed utilizes two open-source simulators, OpenDSS to simulate the electric distribution system and the microscopic traffic simulator SUMO to simulate the traffic dynamics. Electric vehicle charging links the electric distribution system and the transportation network models at vehicle locations determined using publicly available parcel data. Leveraging high-fidelity synthetic electric distribution system data from the SMART-DS project and transportation system data from OpenStreetMap, this testbed models the city of Greensboro, NC down to the household level. Moreover, the methodology and the supporting scripts released with the testbed allow adaption to other areas where high-fidelity geolocated OpenDSS datasets are available. After describing the components and usage of the testbed, we exemplify applications enabled by the testbed via two scenarios modeling the extreme stresses encountered during evacuations.
\end{abstract}


\section{Introduction}
\IEEEPARstart{E}{lectric} vehicles are becoming increasingly popular, with estimates that as much as $15$\% of the passenger vehicle fleet in the US will be electrified by 2030~\cite{usdrive2019}. While the electrification of the vehicle fleet brings substantial benefits, such as reduced pollution and noise, electric vehicles (EVs) will also increase the load on the power grid. Most previous research has focused on independently modeling and analyzing either the transportation network or the power grid with, at best, a significantly simplified model of the other infrastructure. The rapid growth in electric vehicles necessitates joint analyses that couple high-fidelity models of electric distribution systems and transportation networks. 

High-fidelity simulations can accurately assess the impacts of electric vehicles during normal operations of day-to-day charging and travel~\cite{calearo2019,yu2022,rahman2022}. Moreover, high-fidelity simulations are crucial for modeling extreme events such as evacuations that heavily stress both the transportation network (due to traffic congestion) and the power grid (due to the need for rapid and widespread charging prior to evacuating). Since all the traffic is heading in the same direction during an evacuation, standard macroscopic models for traffic become invalid~\cite{nilsson2022evacation}. Furthermore, previously proposed evacuation planning methods~\cite{vanhentenryck2020a, vanhentenryck2020b} may be inapplicable to systems with high penetrations of electric vehicles. These methods focus on scheduling departure routes and times to mitigate traffic congestion during evacuations. Without considering an electric grid model, the charging schedules needed to support the evacuation may overload the electric distribution infrastructure, potentially interrupting the evacuation's progress~\cite{feng2020}. Moreover, charging station infrastructure may be inadequate to support these evacuations~\cite{adderly2018,macdonald2021}.

To enable analyses of both normal conditions and extreme events, this paper presents a testbed for high-fidelity simulations of the charging and movement of electric vehicles within a city. The testbed links together two different simulators, the microscopic traffic simulator SUMO~\cite{SUMO2018} and the power grid simulator OpenDSS~\cite{OPENDSS}. With traffic models of individual vehicles and electric distribution system models down to individual households, these simulators represent more granular levels of detail than what is typically possible to analyze analytically. Hence, we envision that this testbed will serve as a validation platform for tasks such as identifying relevant modeling approximations and assessing control strategies developed using less granular models. Similar testbeds for pure transportation applications have previously been proposed, e.g.,~\cite{turinsumo, lust}. Likewise, synthetic power grid datasets have recently been developed for both transmission systems~\cite{thiam2016,huang2018,fioretto2020,birchfield2017a,birchfield2017b} and distribution systems~\cite{palmintier2021}. As we will discuss more below, our testbed builds on one such electric distribution test system from the SMART-DS project~\cite{palmintier2021,smartds}. 

Much of the previous research on coupled power and transportation systems considers long-term analyses, especially regarding network expansion planning~\cite{xie2020,gan2020}. Some studies, such as~\cite{lai2020}, explore the interactions between these two systems in the medium- and short-term time scales with applications in charging management and vehicle routing. Additionally,~\cite{shuai2020} proposes a method for co-simulation of power grid, traffic, and information networks via data interaction and synchronization between various simulation tools (some of which will be also used in this paper). In-depth reviews of the modeling, interdependence, and applications of these coupled system can be found in~\cite{zheng2019} and~\cite{wei2019}.

To the best of our knowledge, none of this existing literature develops an integrated open-source tool for microscopic traffic simulation and high-fidelity electric distribution system simulation of a geographically comprehensive area.
The main contribution of this paper is thus the description of a ready-to-use testbed for the detailed simulations of the charging and traffic operations of electric vehicles in a moderate-size urban area. We also note that a secondary contribution of this paper is an illustration of how publicly available data from different sources can be combined together to construct testbeds like the one presented in this paper. 

The remainder of this paper is organized as follows. In Section~\ref{sec:components}, we describe the role of the testbed's components and how the data for these components were obtained. In Section~\ref{sec:usage}, we give a high-level explanation for how the testbed can be used and the results that can be produced. For a detailed user manual for the testbed, we refer to the documentation available at \texttt{\url{https://github.com/GreenEVT/GreenEVT}}. In Section~\ref{sec:scanarios}, we illustrate applications of the testbed via two evacuation scenarios that heavily stress both the power grid and the transportation system.

\section{Testbed components}\label{sec:components}

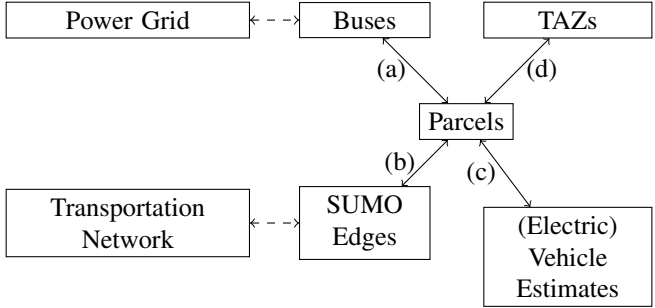
\begin{figure}
    \centering
  \begin{tikzpicture}[scale=0.9]
    \node[draw, text width=3cm, text centered] (tnet) at (0,-0.5) {Transportation\\Network};
    \node[draw, text width=3cm, text centered] (pg) at (0,2.5) {Power Grid};
    \node[draw, text width=1.5cm, text centered] (buses) at (3.5,2.5) {Buses};
    \node[draw, text width=1.5cm, text centered] (edges) at (3.5,-0.5) {SUMO Edges};
    \node[draw] (parcels) at (5,1) {Parcels};
    \node[draw, text width=2cm, text centered] (taz) at (6.5,2.5) {TAZs};
    \node[draw, text width=2cm, text centered ] (veh) at (6.5,-1) {(Electric) Vehicle\\Estimates};

    \draw[<->, dashed] (tnet) --(edges);
    \draw[<->, dashed] (pg) --(buses);
    \draw[<->] (buses) -- node[left] {(a)} (parcels);
    \draw[<->] (edges) -- node[left] {(b)} (parcels);
    \draw[<->] (taz) -- node[right] {(d)} (parcels);
    \draw[<->] (veh) -- node[left] {(c)} (parcels);

    \end{tikzpicture}
    \caption{Conceptual figure of how the different data sources in the testbed are connected. Dashed links represent connections that are already in the respective source data, while the solid links were created through data processing scripts that come with the testbed.}
    \label{fig:components}
    \vspace{-1em}
\end{figure}

In this section, we describe the major components of the testbed and their linkages, which are summarized in Figure~\ref{fig:components}.

\subsection{Power Grid}

The key capability of the testbed is the joint simulation of the transportation and electric distribution networks for the city of Greensboro, NC. To accomplish this, the testbed requires granular and high-fidelity electric distribution data for this city. For this purpose, we use one of NREL’s SMART-DS (Synthetic Models for Advanced, Realistic Testing: Distribution systems and Scenarios) datasets \cite{palmintier2021,smartds}. These datasets are large-scale, realistic-but-not-real electric distribution models that capture electrical connections at all levels of distribution systems down to individual households. The datasets are created using information on actual buildings in combination with synthetic loads that have gone through extensive validation to closely match the behavior of real-life distribution systems.

Electric distribution datasets describe components such as generators and loads attached to buses. The buses are connected by lines and transformers to form a network that is typically operated in a radial (tree-like) structure. For the SMART-DS Greensboro dataset used in this testbed, a single bus representing the transmission system is connected to several different substations through subtransmission lines. Then, after each substation, the network is further subdivided into separate components called feeders, which go all the way down to individual consumers. Transformers step down the voltage at each level of the network. The network is represented by an unbalanced three-phase power flow model which relates the voltages and power injections at each bus and power flows through the lines and transformers.

Each consumer has their own load which we aim to serve without overloading the network's components. The voltage and current limits of each component as well as the consumer loads are provided by the SMART-DS dataset. Slightly overloading some components for a brief period of time might not result in severe damage but repetitive small overloads contribute to loss-of-life for the components. Large overloads can trigger protection mechanisms to avoid severe damage to components, which could leave large numbers of customers without electric service.

\begin{figure}
    \centering
    \includegraphics[width=0.48\textwidth]{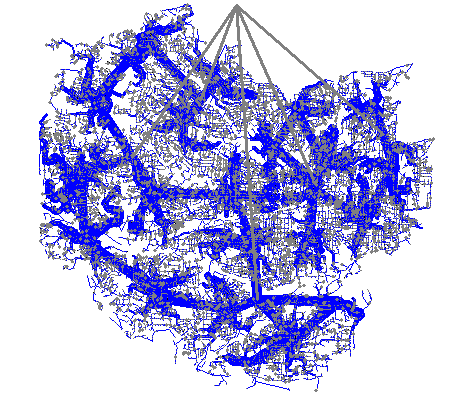}
    \caption{The urban-suburban distribution network of the SMART-DS Greensboro dataset plotted using OpenDSS.}
    \label{fig:power_network}
    \vspace{-1em}
\end{figure}

The full SMART-DS Greensboro dataset contains three separate regions, denoted as ``industrial'', ``rural'', and ``urban-suburban''. In this testbed, we simulate the urban-suburban region shown in Figure~\ref{fig:power_network}, which contains 21 substations, 61 feeders, 154,241 buses and 218,166 total devices. The dataset also includes both peak planning loads and yearly timeseries loads. For the purposes of the testbed, we focus on the peak planning loads. The total active peak load for the urban-suburban region of Greensboro is 612.7~MW. To model electric vehicle charging, we recursively modify individual loads depending on the electric vehicles' placements contained in the testbed's database and vehicle charging schedules informed by the traffic simulations. By default, we assume a constant charging rate of 7.2~kW for each electric vehicle to match the most commonly used EV Level 2 charging equipment \cite{doe}.

To compile the electric network and run power flow simulations, we use EPRI’s Open Distribution System Simulator, OpenDSS~\cite{OPENDSS}.\footnote{\url{https://sourceforge.net/projects/electricdss}} Given a specific load profile, these power flow simulations compute the amount of power and current flowing through each component in the system as well as the voltages at each bus using  a Newton-Raphson method. These results thus indicate which components are operating at or beyond their specified limits. Following OpenDSS's default behavior, we allow the simulator to perform transformer tap changes, switch capacitors, and regulate voltages to decrease the number of components operating beyond these limits. By using the COM interface in an external programming language, one can also use the simulator to export network data, modify system components, execute custom time simulations, and print detailed solution reports. The details for each OpenDSS network component, including load models, transformer settings, and voltage bases are defined in the SMART-DS dataset.

\begin{figure}
    \centering
    \includegraphics[width=0.48\textwidth]{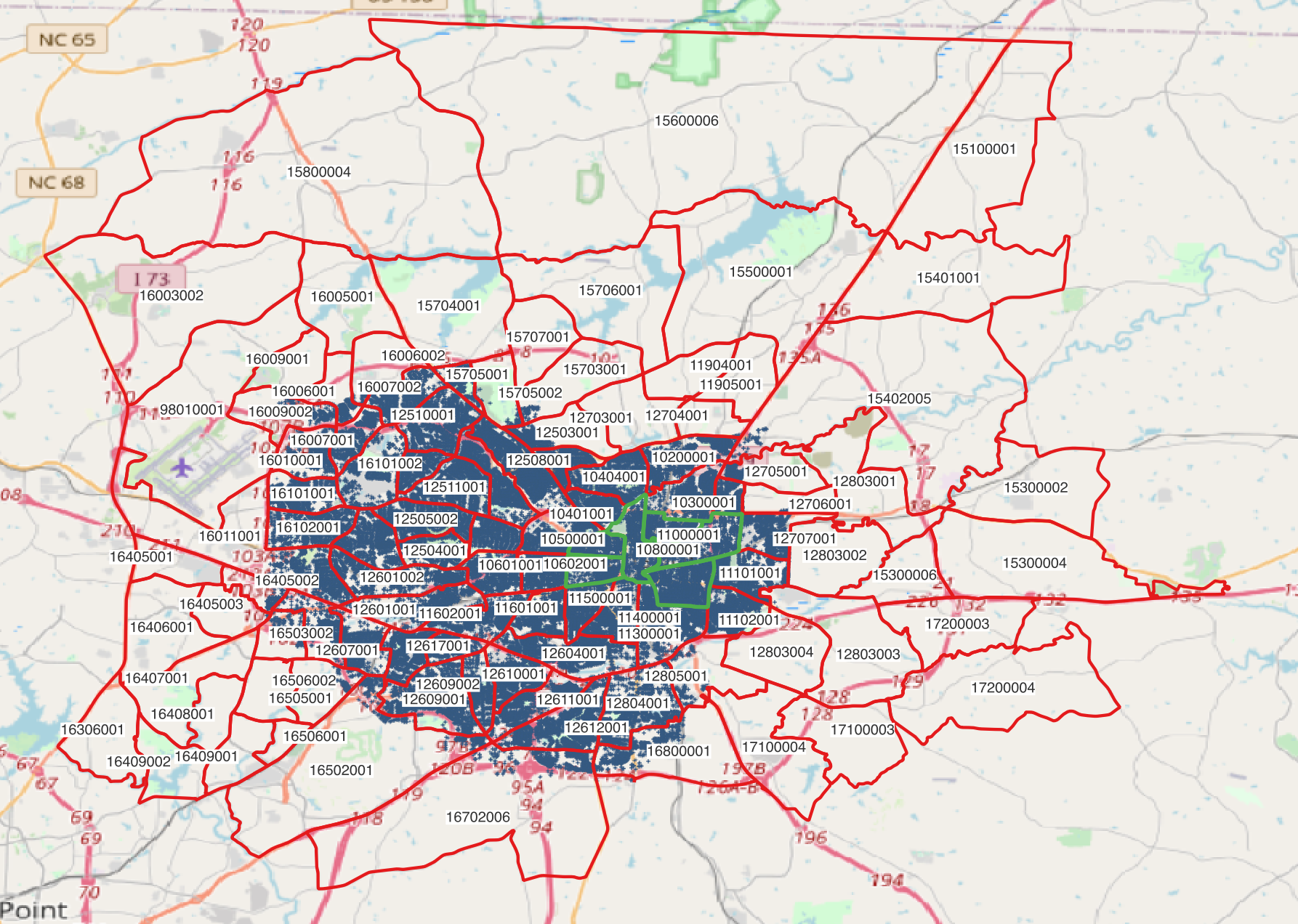}
    \caption{The TAZs in the area of interest. In total, we have selected 106 TAZs, covering the city center's central area and its surroundings. The colored region contains the buses in the power grid network. The green outlined area denotes the six TAZs we are using in the small evacuation scenario described in Section~\ref{sec:scanarios}.}
    \label{fig:transportation_network_withtaz}
\end{figure}

To connect the buses in the electric distribution network, whose geographical position is shown in Figure~\ref{fig:transportation_network_withtaz}, with other components in the testbed, we leverage the geographic coordinates of each bus in the power network dataset to connect each parcel (i.e., an individual plot of land or real estate property) in the area served by the electric distribution system to the closest bus. This is schematically indicated by link (a) in Figure~\ref{fig:components}.

\subsection{Transportation Network}

\begin{figure}
    \centering
    \includegraphics[width=0.48\textwidth]{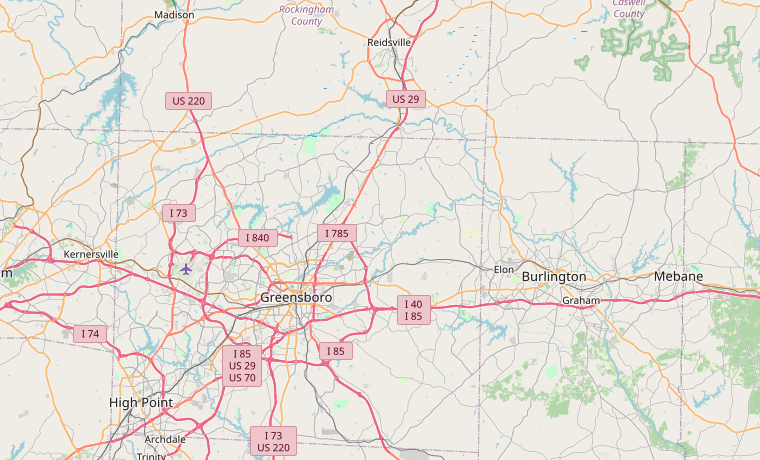}
    \caption{The area of the transportation network considered. Picture from OpenStreetMap.org.}
    \label{fig:transportation_network}
    \vspace{-1em}
\end{figure}

\begin{figure}
    \centering
    \includegraphics[width=0.48\textwidth]{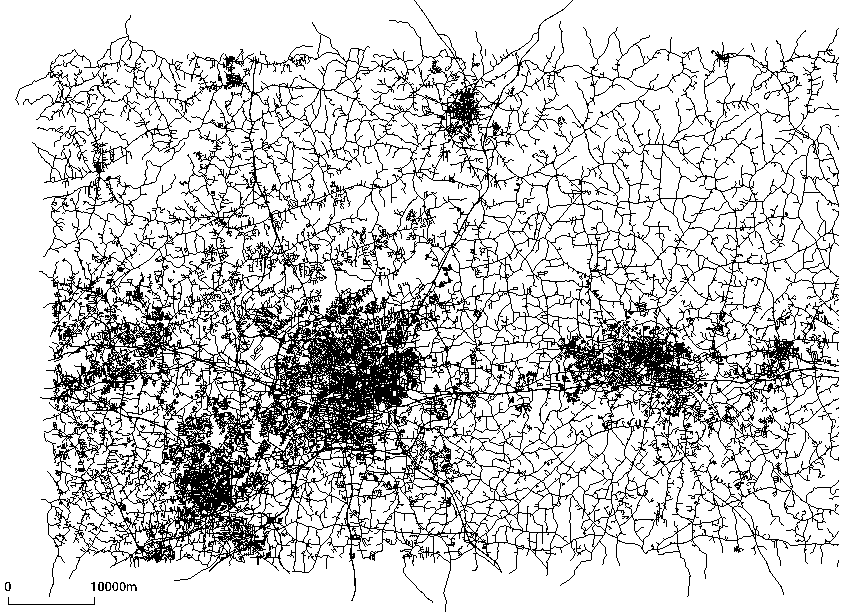}
    \caption{The road network imported in SUMO.}
    \label{fig:SUMO_Network}
    \vspace{-1em}
\end{figure}

\begin{table}[]
    \centering
    \caption{Properties of the transportation network.}
    \begin{tabular}{ll}
    \hline
        Total road length & 28216 km \\
        Total lane length & 30625 km \\
        No of signalized junctions & 914 \\ 
        No of SUMO nodes & 50109  \\
        No of SUMO edges & 118289  \\ \hline
    \end{tabular}
    \label{tab:SUMOnetworkstatistics}
    \vspace{-1em}
\end{table}

The transportation network in the testbed was imported from OpenStreetMap using the tool osmWebWizard.\footnote{\url{https://sumo.dlr.de/docs/Tools/Import/OSM.html}} The map covers a rectangular area\footnote{The area covered is between latitude $35.894145$ longitude $-80.198958$ (N $35^\circ53'38.922''$, W$80^\circ11'56.284''$) and  latitude $36.410277$ longitude $-79.157255$ (N $36^\circ24'36.9972"$, W$79^\circ9' 26.118"$)} as shown in Figure~\ref{fig:transportation_network}. The imported SUMO network is shown in Figure~\ref{fig:SUMO_Network}. Table~\ref{tab:SUMOnetworkstatistics} overviews the transportation network properties, where an edge in SUMO represents a road or a part of a road. Leveraging functions in the SUMO Python Library, Sumolib, we located the closest edge to each parcel and saved the corresponding identifier for the SUMO edge in the testbed's parcel table, i.e., made linkage (b) in Figure~\ref{fig:components}. Using the number of vehicles associated with each parcel along with this linkage, the testbed inserts the vehicles into the SUMO simulator once a vehicle is scheduled to depart.

\subsection{Parcel Data and Vehicle Estimation}
As shown in Figure~\ref{fig:components}, the parcel data plays a central role in the testbed. Using public data sources about Guildford Country, where Greensboro is located, from the state of North Carolina,\footnote{\url{https://www.nconemap.gov/pages/parcels}} the testbed incorporates data on each parcel's location and usage (e.g., designation as an apartment complex, an office building, a single-family residential building, etc.). Depending on its usage, each parcel may correspond to either a single building or a group of buildings. In total, there are $136,655$ parcels in the area covered. 

The testbed also links each parcel to its Traffic Analysis Zone (TAZ), i.e., link (d) in Figure~\ref{fig:components}. Traffic analysis zones are commonly used in traffic planning as a suitable way to divide the city into geographically smaller regions. The testbed's traffic analysis zones were imported from open data provided by the U.S. Census Bureau.\footnote{\url{https://www2.census.gov/geo/tiger/TIGER2010/TAZ/2010/}} All the TAZs in the area we are studying are shown in Figure~\ref{fig:transportation_network_withtaz}.

We estimate the number of vehicles for each parcel based on the parcel's usage specification as shown in Table~\ref{tab:parcels}. A manual estimate with visual support from Google Maps and Google Streetview was performed for parcels that do not fall into any of the categories mentioned in Table~\ref{tab:parcels}. The web interface for manual counting is also provided with the testbed in the file \filepath{scripts/manualvehiclecount.php}. The combination of vehicle estimation by category and manual counting creates the link labeled (c) in Figure~\ref{fig:components} with respect to the total number of vehicles (both electric and conventional). Note that the vehicle estimates consider one vehicle per household in accordance with an evacuation scenario. For more realistic day-to-day usage scenarios, the numbers should be increased appropriately. The testbed easily accommodates such parameter modifications via updating the estimate for each category in the database and scaling the manual vehicle count in an appropriate manner. Since many of the parcels with a large number of manually counted vehicles are apartment complexes, upscaling counts by an appropriate factor would be sufficient for this class of parcels.  
\begin{table}
    \centering

    \caption{Estimated number of vehicles for selected parcel categories}
    \resizebox{\columnwidth}{!}{%
    \begin{tabular}{llll}
    \multirow{2}{*}{Category} & \multirow{ 2}{*}{Subcategory} & Vehicle & \multirow{ 2}{*}{Comment} \\
    & & Estimate & \\ \hline 
APART&	07-APT<5 UNITS	&4 &	Apartments with less than 5 units\\
RESIDENTIAL&	07-APT<5 UNITS&	4	& Multi-family houses with less than five units\\
GOV OWNED&	07-APT<5 UNITS&	4	&Apartments with less than 5 units \\
APART&	041-TOWNHOME &	4	& Only 1 in Guilford, has 4 units\\
COMM&	07-APT<5 UNITS&4	& Apartments with less than 5 units\\
COMM&	09-TWNHSEAPT&	4&	Only 1 in Guilford, has 4 units\\
RESIDENTIAL&	08-DUPLEX/TRIPLEX&	3	&Duplexes and Triplexes (chose 3 to be conservative) \\
MULTI-FAMILY<4 &	01-SFR &	3	&Multi-family houses with less than four units \\
MULTI-FAMILY<4	&08-DUPLEX/TRIPLEX&	3&	Duplexes and Triplexes (chose 3 to be conservative)\\
GOV OWNED&	08-DUPLEX/TRIPLEX&	3&	Duplexes and Triplexes (chose 3 to be conservative)\\
MULTI-FAMILY<4&	07-APT<5 UNITS&	3&	Multi-Family Homes\\
MULTI-FAMILY<4&	0&	3&	Multi-family residences with less than 4 units\\
OFFICE&	08-DUPLEX/TRIPLEX&	3&	Duplexes and Triplexes (chose 3 to be conservative)\\
COMM&	08-DUPLEX/TRIPLEX&	3&	Duplexes and Triplexes (chose 3 to be conservative)\\
APART&	08-DUPLEX/TRIPLEX&	3&	Duplexes and Triplexes (chose 3 to be conservative)\\
IND&	08-DUPLEX/TRIPLEX&	3&	Duplexes and Triplexes (chose 3 to be conservative)\\
RESIDENTIAL&	01-SFR&	1&	SFR = Single Family Residential\\
CONDO&	04-CONDO&	1&	Residential Condos (each condo has a separate parcel)\\
TOWNHOUSE&	041-TOWNHOME&	1&	Townhomes \\
RESIDENTIAL&	09-TWNHSEAPT&	1&	Townhouse\\
AGRI/HORT&	01-SFR&	1&	Houses with Farmland\\
RESIDENTIAL&	02-MANUFHM&	1&Houses (manufactured houses)\\
COMM&	01-SFR&	1&	Single Family Residence, near commercial district\\
GOV OWNED&	01-SFR&	1& Single Family Residences \\
DEVEL. RESTRICT. &	01-SFR&	1&	Single Family Residences \\
IND&	01-SFR&	1&	Single Family Residences \\
INSTITUTIONAL&	01-SFR&	1&	Single Family Residences \\
SINGLE WIDE MH&	0&	1&	Motorhomes\\
ASSIST LIV/SKILLCARE&	01-SFR&	1&	Single Family Residences \\
OFFICE&	01-SFR&	1&	Single Family Residences \\
APART&	01-SFR&	1&	Single Family Residences \\
AIRPORT&	01-SFR&	1&	Single Family Residences \\
VACANT&	01-SFR&	1&	Single Family Residences \\
MFG HOM&	02-MANUFHM&	1&	Manufactured Homes\\
TOWNHOUSE&	04-CONDO&	1	&1 Condo per parcel\\
SCHOOL/COLL/UNIV&	01-SFR&	1&	Single Family Residences \\
RESIDENTIAL&	05-PATIOHM&	1&	Single Family Patio Homes\\
TWINHOME&	012-TWIN HOME&	1&	Twin Homes\\
COMM&	02-MANUFHM&	1	&Manufactured Homes\\
INSTITUTIONAL	&02-MANUFHM&	1	&Manufactured Homes\\
LEASED&	01-SFR&	1&	Single Family Residences \\
IND	&02-MANUFHM	&1&	Manufactured Homes \\ \hline
    \end{tabular}}
    \label{tab:parcels}
\end{table}

\begin{table}
 \caption{Number of vehicles in each TAZ}
    \label{tab:my_label}
\centering
    \begin{tabular}{ccc}
No. &  TAZ & No. vehicles  \\ \hline 
1 & 10100001 &	678  \\
2 & 10200001 &	1781 \\
3 & 10300001 &	973 \\
4 & 10401001 &	781 \\
5 & 10403001 &	796 \\
6 & 10404001 &	1087 \\
7 & 10500001 &	1164 \\
8 & 10601001 &	1463  \\
9 & 10602001 &	2645 \\
10 & 10701001 &	1131 \\
11 & 10702001 &	1192 \\
12 & 10800001 &	1281 \\
13 & 10900001 &	993 \\
14 & 11000001 &	813 \\
15 & 11101001 &	2148 \\
16 & 11102001 &	1309 \\
17 & 11200001 &	2051 \\
18 & 11300001 &	1374 \\
19 & 11400001 &	1264 \\
20 & 11500001 &	1294 \\
21 & 11601001 &	1076 \\
22 & 11602001 &	993 \\
23 & 11904001 &	2024 \\
24 & 11905001 &	1843 \\
25 & 12503001 &	1871 \\
26 & 12504001 &	1103 \\
27 & 12505001 &	752 \\
28 & 12505002 &	1161 \\
29 & 12508001 &	1668 \\
30 & 12509001 &	1088 \\
31 & 12510001 &	1438 \\
32 & 12511001 &	1378 \\
33 & 12511002 &	708 \\
34 & 12601001 &	1816 \\
35 & 12601002 &	972 \\
36 & 12604001 &	1961 \\
37 & 12607001 &	995 \\
38 & 12608001 &	1241 \\
39 & 12609001 &	2473 \\
40 & 12609002 &	941 \\
41 & 12610001 &	1133 \\
42 & 12611001 &	1140 \\
43 & 12612001 &	2618 \\
44 & 12617001 &	971 \\
45 & 12703001 &	1921 \\
46 & 12704001 &	1310 \\
47 & 12705001 &	1444 \\
48 & 12706001 &	1429 \\
49 & 12707001 &	1132 \\
50 & 12803001 &	469  \\
51 & 12803002 &	865 \\
52 & 12803003 &	1478 \\
53 & 12803004 &	437 \\ \hline
\end{tabular}
    \begin{tabular}{ccc}
      No. &  TAZ & No. vehicles  \\ \hline 
54 & 12804001 &	1583 \\
55 & 12805001 &	836 \\
56 & 15100001 &	891 \\
57 & 15300002 &	918 \\
58 & 15300004 &	831 \\
59 & 15300006 &	295 \\
60 & 15401001 &	1474 \\
61 & 15402005 &	2686 \\
62 & 15500001 &	2567 \\
63 & 15600006 &	3865 \\
64 & 15703001 &	2622 \\
65 & 15704001 &	2455 \\
66 & 15705001 &	948 \\
67 & 15705002 &	438 \\
68 & 15706001 &	2455 \\
69 & 15707001 &	2566 \\
70 & 15800004 &	3072 \\
71 & 16003002 &	1726 \\
72 & 16005001 &	780 \\
73 & 16006001 &	488 \\
74 & 16006002 &	679 \\
75 & 16007001 &	1023 \\
76 & 16007002 &	998 \\
77 & 16009001 &	916 \\
78 & 16009002 &	567 \\
79 & 16010001 &	1218 \\
80 & 16011001 &	2162 \\
81 & 16101001 &	1150 \\
82 & 16101002 &	482 \\
83 & 16102001 &	1336 \\
84 & 16103001 &	1954 \\
85 & 16306001 &	2522 \\
86 & 16405001 &	10 \\
87 & 16405002 &	923 \\
88 & 16405003 &	34 \\
89 & 16406001 &	1632 \\
90 & 16407001 &	1835 \\
91 & 16408001 &	803 \\
92 & 16409001 &	362 \\
93 & 16409002 &	626 \\
94 & 16502001 &	2365 \\
95 & 16503001 &	913 \\
96 & 16503002 &	1223 \\ 
97 & 16505001 &	1621 \\
98 & 16506001 &	488 \\ 
99 & 16506002 &	1363 \\
100 & 16702006 &	2346 \\
101 & 16800001 &	814 \\
102 & 17100003 &	541 \\
103 & 17100004 &	599 \\
104 & 17200003 &	654 \\
105 & 17200004 &	393 \\
106 & 98010001 &	2931 \\ \hline
    \end{tabular}
   
\end{table}

\subsection{Prediction on Electric Vehicle Adoption}
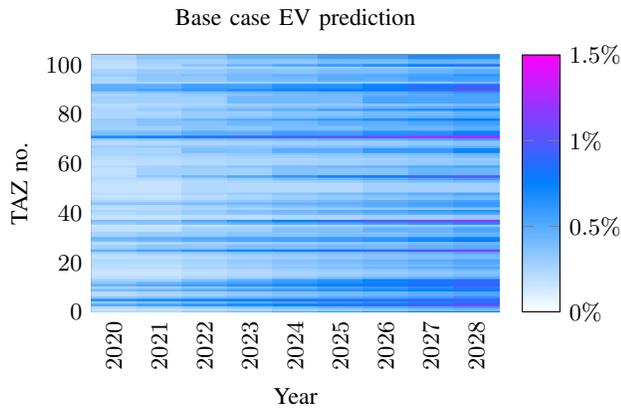
\begin{figure}
\centering
\begin{tikzpicture}
\pgfkeys{/pgf/number format/.cd,1000 sep={}}
\pgfplotsset{scaled ticks = false}
\begin{axis}[view={0}{90}, width=7cm, height=5cm, colorbar, 
  colormap/cool, ylabel={\small TAZ no.}, xlabel={Year}, xtick=data, xticklabel style = {rotate=90,anchor=east}, ymin=0, ymax=104, ylabel near ticks, ytick style={draw=none}, title={Base case EV prediction},  colorbar style={   yticklabel={$\pgfmathprintnumber\tick\%$}},
  xmin={2019.5}, xmax={2028.5}, point meta min=0.0, point meta max=1.5
]
\addplot [matrix plot*, point meta=explicit] 
table [
                    x index=0,
                    y index=1,
                    meta expr={100*\thisrowno{2}},
                ] {data/base.dat};
\end{axis}
\end{tikzpicture}
\caption{Percentage of electric vehicles in each TAZ in the base case prediction.}
\label{fig:EVpredictionbase}
\end{figure}

\begin{figure}[t!]
\centering
\begin{tikzpicture}
\pgfkeys{/pgf/number format/.cd,1000 sep={}}

\begin{axis}[view={0}{90}, width=7cm, height=5cm,  colorbar, colorbar style={        yticklabel={$\pgfmathprintnumber\tick\%$}},
  colormap/cool, ylabel={\small TAZ no.}, xlabel={Year}, xtick=data, xticklabel style = {rotate=90,anchor=east}, ymin=0, ymax=104, title={High case EV prediction}
, xticklabel style = {font=\tiny}, xmin={2019.5}, xmax={2050.5}, ylabel near ticks,
 point meta min=0.0,
point meta max=70
]
\addplot[matrix plot*, point meta=explicit ] table [
                    x index=0,
                    y index=1,
                    meta expr={100*\thisrowno{2}},
                ] {data/high.dat};
\end{axis}
\end{tikzpicture}
\caption{Percentage of electric vehicles in each TAZ in the high case prediction.}
\label{fig:EVpredictionhigh}
\vspace{-1em}
\end{figure}
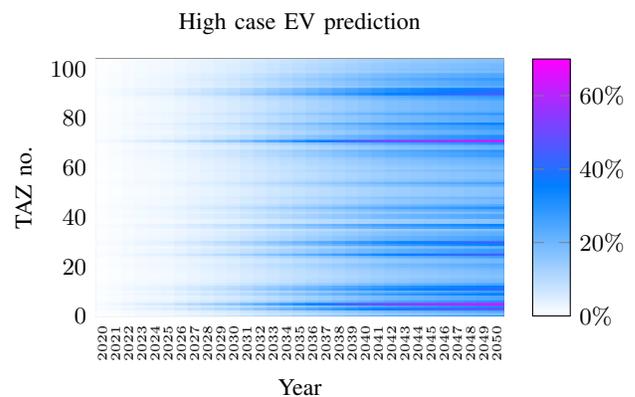

To predict the number of electric vehicles, we consider three main cases: one \emph{base} case and two cases with higher EV penetration rates, which we denote as the \emph{medium} and \emph{high} cases. 
The base case directly uses historical data along with a prediction of future sales that is based on past electric vehicle sales backdated by the average vehicle age.
This prediction is likely to be a conservative estimate since it does not incorporate increasing electric vehicle adoption rates. The medium and high cases are based on projections from USDRIVE~\cite{usdrive2019} that account for increasing interest in buying electric vehicles. Figures~\ref{fig:EVpredictionbase}--\ref{fig:EVpredictionhigh} show the percentages of electric vehicles in each TAZ for each case (base, medium, and high).
Additionally, to perform a sensitivity analysis on power grid impacts, we consider an \emph{extreme} case in one of the demonstrations in Section~\ref{sec:scanarios}. This extreme case is the same as the \emph{high} case but with additional EVs assigned at random to reach an 80\% EV penetration rate.

\subsubsection{Base Case}

To generate a reasonable distribution of their current status, we used electric vehicle registration data from Guilford County in 2019.\footnote{\url{https://www.ncdot.gov/initiatives-policies/environmental/climate-change/Pages/zev-registration-data.aspx}} Electric vehicles were then distributed among the census tracts proportionally to the number of households and the median income in each census tract. In this way, census tracts with more households and higher incomes are assigned higher fractions of electric vehicles.

For the base case, we used data on the average vehicle age based on household income from the National Household Travel Survey.\footnote{\url{https://nhts.ornl.gov}} The data are extrapolated to the year 2028. This extrapolation is then combined with the mean household income in each census tract\footnote{\url{https://data.census.gov/cedsci/table?q=income&g=0500000US37081.140000&tid=ACSST5Y2018.S1901&moe=false&hidePreview=true}} to find the average vehicle age for each census tract between the year 2020 and the year 2028. We assume here that the mean household income will not change significantly during the prediction period. 

Once we know the expected average age of vehicles in each census tract, the number of new electric vehicles added yearly to each census tract is estimated by the fraction of electric vehicles that were sold the average-vehicle-age ago. For example, if the average vehicle age is $10$ years, we use vehicle sales from 2015 to predict how many new EVs will appear in the year 2025. This approach captures the fact that census tracts with newer vehicles, on average, are more likely to consider buying electric vehicles.

To find the historical market penetration rate, we divided the number of EVs sold in the US each year\footnote{See \url{https://electricdrive.org/index.php?ht=d/sp/i/20952/pid/20952}, \url{https://insideevs.com/news/344006/monthly-plug-in-report-card-archive/}, and \url{https://afdc.energy.gov/data/10567}} by the total vehicle sales for the same year.\footnote{\url{https://fred.stlouisfed.org/series/TOTALSA#0}} To adapt those fractions to Guildford country, we utilized data from 2019 on the number of registrations of electric vehicles in North Carolina\footnote{\url{https://afdc.energy.gov/data/10962}} and the total number of electric vehicles sold in Guildford county.\footnote{\url{https://www.ncdot.gov/initiatives-policies/environmental/climate-change/Pages/zev-registration-data.aspx}}

\begin{figure}
\centering
\begin{tikzpicture}
\pgfkeys{/pgf/number format/.cd,1000 sep={}}

\begin{axis}[view={0}{90}, width=7cm, height=5cm,   colorbar, colorbar style={        yticklabel={$\pgfmathprintnumber\tick\%$}},
  colormap/cool, ylabel={\small TAZ no.}, xlabel={Year}, xtick=data, xticklabel style = {rotate=90,anchor=east}, ymin=0, ymax=104, title={Medium case EV prediction},  xticklabel style = {font=\tiny}, xmin={2019.5}, xmax={2050.5}, ylabel near ticks,
   point meta min=0.0,
point meta max=70
]

\addplot [matrix plot*,point meta=explicit] table [
                    x index=0,
                    y index=1,
                    meta expr={100*\thisrowno{2}},
                ] {data/medium.dat};
\end{axis}
\end{tikzpicture}
\caption{Percentage of electric vehicles in each TAZ in the medium case prediction.}
\label{fig:EVpredictionmedium}

\end{figure}
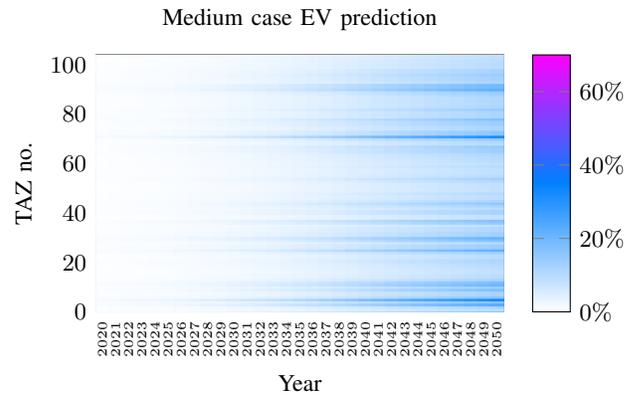

Since TAZs are more granular than census tracts, the number of electric vehicles were split equally among the census tracts when there are several TAZs within a census tract. For a few census tracts, parts of the tracts were located outside the area covered. In this case, the prediction of the number of electric vehicles was scaled down in proportion to the land area within the TAZ.

To get the percentage of electric vehicles within each TAZ, we adopted the projected total fleet size (including both electric and conventional vehicles) for the United States from 2020--2050\footnote{\url{https://www.eia.gov/outlooks/aeo/data/browser/#/}
} to Guildford country by utilizing the percentage of the total number of vehicles sold in Guildford county\footnote{\url{https://www.ncdot.gov/initiatives-policies/environmental/climate-change/Pages/zev-registration-data.aspx}} for 2019. By assuming that that Guildford country accounts for a constant proportion of all vehicles in the United States, we then projected values for the total number of vehicles in Guildford from 2020 to 2050. This total fleet size was split in proportion to the number of households in each census tract. The total fleet sizes were adjusted to TAZs in the same manner as the EVs, finally enabling calculation of the percentage of electric vehicles in each TAZ.

\subsubsection{Medium and High Cases}

Predictions from USDRIVE~\cite{usdrive2019} were used to construct the medium and high cases. Historical data of the electric vehicle sales per state\footnote{\url{https://afdc.energy.gov/data/10962}} and per county in North Carolina\footnote{\url{https://www.ncdot.gov/initiatives-policies/environmental/climate-change/Pages/zev-registration-data.aspx}} were used to apply the USDRIVE predictions to Guilford county. To spatially distribute the predicted number of electric vehicles in each census tract, we assumed that the electric vehicles will be distributed proportionally to both the current number of households in each census tract and the median household income in each census tract. The predicted numbers of electric vehicles for each census tract were then converted to predicted numbers of electric vehicles for each TAZ by following the same methodology as for the base case scenario. The same methodology as for the base case was also used to convert the absolute number of electric vehicles to the fraction of electric vehicles.

Note that we only use these predictions of vehicle electrification trends to determine the \emph{fraction} of electric vehicles in each TAZ. In the testbed, the total number of vehicles remains constant every year since the testbed does not include further housing or other infrastructure development, such as the expansion of the power grid. In this way, we maintain consistency between the capacities of the electric and transportation infrastructure and the vehicle travel and charging demands, thus avoiding the need to consider infrastructure expansion over time.

Interpreting the fraction of electric vehicles within a TAZ as the probability that each vehicle associated with a parcel is electric, the last part of the link (c) in Figure~\ref{fig:components} is completed, and hence all data needed to perform simulations in the testbed are in place. In the next section, we will describe the pipeline for running simulations in the testbed.

\section{Testbed Usage}\label{sec:usage}

The testbed consists of several modules that can be run either jointly or separately, given that the data each module depends on has been generated at some earlier time. The testbed is applied to analyze \emph{scenarios}, each of which has the following properties:
\begin{itemize}
    \item \emph{Fraction of electric vehicles:} A scenario can either have a fixed fraction of electric vehicles, i.e., every vehicle has a fixed probability of being electric, or a prediction using the approach described in the previous section.
    \item \emph{Departure time}: Each vehicle departs at a specified time.
    \item \emph{Charging schedule}: Each electric vehicle starts and stops charging at specified times, using a constant charging rate.
\end{itemize}
The evacuations that we consider as illustrative applications for the testbed additionally require specifying:
\begin{itemize}
    \item The TAZs that will be evacuated. 
    \item The safe node for the evacuation, i.e., where the vehicles should go once they depart. Since the traffic simulation only covers the area surrounding Greensboro, this safe node corresponds to a point on a freeway at the perimeter of the city, as is typically the case in evacuation planning problems~\cite{vanhentenryck2020a,vanhentenryck2020b}.
\end{itemize}

\subsection{Data Pre-processing Pipeline}
To facilitate the usage of this testbed methodology for other regions, the pre-processing scripts we developed for Greensboro are included in the testbed. After obtaining the SUMO network, OpenDSS network, and parcel data, the pipeline for linking the different datasets can be done with the code \filepath{scripts/data\_preprocessing.py}. The pre-processing script creates a common ground through an SQLite database that both the traffic simulator SUMO and the power simulator OpenDSS draw upon while providing relative independence in terms of running different kinds of simulations, as they both read from the database while not interfering with each other. While most aspects of this pre-processing are straightforward to extend to different regions, note that the data will still likely require some manual processing, such as adding vehicle estimates for some categories of parcels, fixing broken roads in the imported SUMO network, etc. 

For the scenarios we demonstrate in this paper, note that all of the pre-processing is already done, and the outcome is stored in the database that comes with the testbed.

\subsection{Overview of the Workflow}
Once the preprocessing is done, the testbed is executed using the Python script \filepath{interactive\_simulator.py}. Before running this script, the parameters listed in Table~\ref{tab:simulator_parameters} need to be specified. The script then displays a menu in the terminal where the user can choose which stages of the workflow the user wants to (re)-run. The different stages are described below:
\begin{table*}
    \centering
    \caption{Parameters to be set in main simulator script}
    \begin{tabular}{ll}
        Parameter &  Description \\ \hline
        scenario\_name & A unique name for each scenario \\
        working\_dir & The location where the output files will be stored \\
        ev\_penetration\_rate & A value between $0$ and $1$ if the scenario has a fixed penetration rate, $-1$ if using any of the predicted rates \\ 
        year\_prediction & The year considered when utilizing the ``base'', ``medium'', or ``high'' cases \\ 
        prediction\_level & Either ``base'', ``medium'', or ``high'' when utilizing one of these cases \\
        load\_per\_charging\_ev & The electric power demand from each electric vehicle when charging \\
        charging\_time & The amount of time before departure that each vehicle should start charging \\
        departure\_window  & The window during which each vehicle will randomly depart \\
        tazs\_to\_evacuate & Evacuation specific parameter: Set of TAZs to be evacuated \\
        evac\_edge & Evacuation specific parameter: The SUMO edge all vehicles head towards \\
        \hline
    \end{tabular}
    \label{tab:simulator_parameters}
\end{table*}

\subsubsection{Copying Files}
The testbed first copies configuration files for the SUMO and OpenDSS simulators onto the specified file path for the current scenario. These configuration files are not scenario dependent, but the user may want to change some of the simulator-specific parameters in these files. Some parameters for the SUMO simulator, such as maximum simulation time, can be changed directly in the generated configuration file for the simulator. 

Each scenario will have a copy of the simulation files for the SUMO configuration, power grid data, traffic flow files, vehicle route files, etc. For example, two simulations of the same region with the same simulated vehicle locations but different electric vehicle penetration rates or evacuation routes will share the same database but have separate working directories and generate result files in different sub-directories.

The SUMO simulator requires a detailed description of each vehicle's path in the traffic network. To obtain such a low-level description, the SUMO tool duarouter\footnote{\url{https://sumo.dlr.de/docs/duarouter.html}} is utilized. 

\subsubsection{Traffic Simulation Configuration}
In this stage, various configurations for the SUMO simulator are generated. First, the vehicle generation step populates the database with vehicles for the scenario and also determines whether each vehicle is electric or conventional. Next, the testbed assigns departure and charging times for each vehicle in the specified scenario. Both of these steps utilize random numbers, so rerunning the testbed with the same scenario may result in updates to these properties for each vehicle. All other steps in the workflow are deterministic and will hence yield the same results if they are rerun.

\subsubsection{Run Simulations}
The next stage runs the SUMO simulator and the OpenDSS simulator with the parameters set in the main script using the previously created configuration files. Each simulator produces separate output files that can be analyzed. Several scripts for analyzing the simulators' outputs are included with the testbed, such as the scripts for generating the plots in the demonstration cases shown in the next section.

\subsection{Architecture}
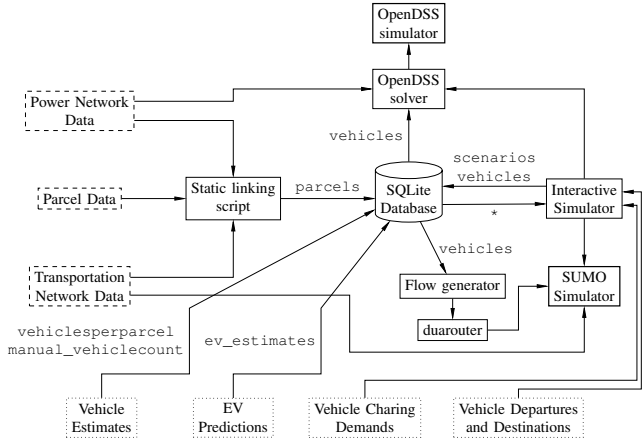
\begin{figure}
    \centering
\resizebox{0.48\textwidth}{!}{
    \begin{tikzpicture}[ database/.style={
      cylinder,
      shape border rotate=90,
      aspect=0.25,
      draw}, 
      >=narrow
 ]

\node[database, align=center] (sqllite) at (0,0) {SQLite\\Database};

\node[draw, align=center,dashed] (powernetwork) at (-7.5,2) {Power Network\\Data};
\node[draw, align=center,dashed] (transportationnetwork) at (-7.5,-2) {Transportation \\ Network Data};
\node[draw, align=center, dashed] (parceldata) at (-7.5,0) {Parcel Data};

\node[draw, align=center] (linking) at (-4,0) {Static linking\\script};

\node[draw, align=center, dotted] (vehicleestmation) at (-7,-5) {Vehicle\\Estimates};
\node[draw, align=center, dotted] (evpred) at (-4,-5) {EV\\Predictions};

\node[draw, align=center, dotted] (charingdemands) at (-1,-5) {Vehicle Charing\\Demands};
\node[draw, align=center, dotted] (vehdep) at (2.5,-5) {Vehicle Departures \\ and Destinations};

\node[draw, align=center] (sim) at (4,0) {Interactive\\Simulator};

\node[draw, align=center, thick] (sumo) at (4,-2) {SUMO\\Simulator};

\node[draw, align=center] (flow) at (1,-2) {Flow generator};
\node[draw, align=center] (router) at (1,-3) {duarouter};

\draw[->] (powernetwork.-10) -| (linking); 
\draw[->] (parceldata) -- (linking); 
\draw[->] (transportationnetwork.10) -| (linking); 
\draw[->] (linking) -- node[above]{\texttt{parcels}} (sqllite);

\node[draw, align=center] (opendssvol) at (0, 2.5) {OpenDSS \\ solver};
\node[draw, align=center, thick] (opendsssim) at (0, 4) {OpenDSS \\ simulator};

\draw[->] (powernetwork.10) -| (-4, 2.5) -- (opendssvol);
\draw[->] (sqllite) -- node[left] {\texttt{vehicles}} (opendssvol);
\draw[->] (opendssvol) -- (opendsssim);

\draw[->] (vehicleestmation) -- +(0,1) -- + (2,1) -- node[left, align=center]{\texttt{vehiclesperparcel}\\\texttt{manual\_vehiclecount}} +(2,2.5) -- (sqllite.-162);

\draw[->] (evpred) -- +(0,1) -- + (2,1) -- node[left, align=center]{\texttt{ev\_estimates}} +(2,2.5) -- (sqllite.-130);

\draw[->] (sim) |- (opendssvol);

\draw[->] (sim.162) --  node[above, align=center]{\texttt{scenarios}\\\texttt{vehicles}}  (sqllite.20);
\draw[->]  (sqllite.-10)  -- node[below, align=center]{\texttt{*}}   (sim.188);

\draw[->] (sqllite) -- node[right] {\texttt{vehicles}} (flow);
\draw[->] (flow) -- (router);
\draw[->] (router) -- +(1.5,0) |- (sumo);

\draw[->] (transportationnetwork.-10) -- +(5,0) |- +(5, -1.3) -|  (sumo);

\draw[->] (sim) -- (sumo);

\draw[->] (charingdemands.90) -- +(0, 0.4) -- +(6.2,0.4) |- (sim.-10);
\draw[->] (vehdep.90) -- +(0, 0.2) -- +(2.8,0.2) |- (sim.10);

\end{tikzpicture}
    }
    \caption{The detailed architecture of the testbed. The dashed boxes are data obtained from exogenous data sources, while the dotted boxes represent data provided by the user. Next to some of the arrows, the figure indicates which table in the database is used or changed by a script. The simulator utilizes several of the generated tables when creating scenarios with vehicles. These dependencies are denoted an asterisk. }
    \label{fig:architecture}
\end{figure}

Figure~\ref{fig:architecture} describes the testbed's architecture. The figure shows when different tables are created or expanded and which component uses each table. Note that the vehicle table contains the vehicles for every scenario, but only the vehicles for the specific scenario are selected by the simulator.

\section{Demonstrations of the Testbed }\label{sec:scanarios}

This section first presents an illustrative application of the testbed via two evacuation scenarios (a small scenario that runs for a few minutes on a personal computer and a large scenario which requires several hours to run the traffic simulator), followed by a discussion of other potential applications.

For both evacuation scenarios, the vehicles evacuate to Route 40 east of Greensboro, and each vehicle will be routed such as it will take its shortest path with respect to travel time as if there were no other vehicles present. For the small scenario, we are using the EV prediction rates for the year 2040, while for the large scenario, we are using the EV prediction rates for the year 2045.

\subsection{Small Evacuation Scenario}
As a simple demonstration scenario that can run quickly on a personal computer, the testbed comes with a scenario where only a small part of the region covered by the testbed is evacuated. The region consists of six TAZs, outlined in green in Figure~\ref{fig:transportation_network_withtaz}. We run several variants of this scenario, considering either the simultaneous departures of all vehicles or vehicle departure times that are spread out randomly over a two-hour window. The scenarios are also run with different penetration rates of electric vehicles to illustrate the evacuation's impacts on the power grid.

\begin{figure}
    \centering
    \input{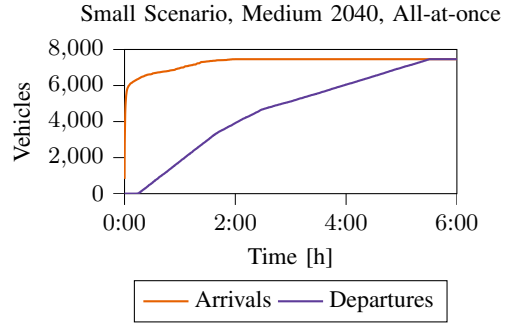}
    \caption{Cumulative departure and arrival curves for the Small Scenario Medium 2040 All-at-once.}
    \label{fig:small_scenario_cumulative_allatonce}
    \vspace{-1em}
\end{figure}

\begin{figure}
    \centering
    \input{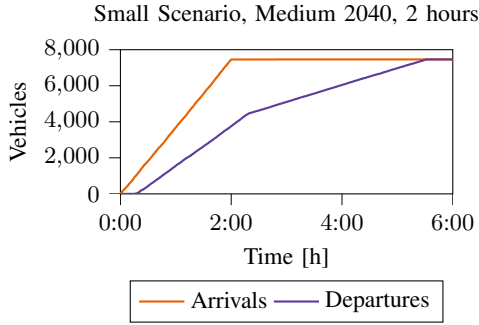}
    \caption{Cumulative departure and arrival curves for the Samll Scenario Medium 2040 with a two-hour departure window.}
    \label{fig:small_scenario_cumulative_2h}
    \vspace{-1em}
\end{figure}

Table~\ref{tab:small_scenario} shows the total number of vehicles and the number of electric vehicles for each scenario. Moreover, the performance of the transportation network is shown. From the performance metrics, we can see that it takes approximately the same time to evacuate regardless if the vehicles are spread out over a two-hour window or not, with the vehicles spending less time waiting in the transportation network when their departure time is spread out. This can further be seen in Figures~\ref{fig:small_scenario_cumulative_allatonce} and~\ref{fig:small_scenario_cumulative_2h}, where we plot the cumulative number of departures and arrivals. In both figures, the curves intersect at almost the same point, but the area between the curves is considerably less for the two-hour departure window scenario, meaning that the vehicles spend less time on the road.

\begin{table*}
\centering
\caption{Statistics regarding the transportation network for the small scenario}
\begin{tabular}{ccccc} 
& Medium 2040 All-at-once & Medium 2040 2 hours &  High 2040 All-at-once & High 2040 2 hours \\ \hline
Number of vehicles & $7461$  & $7461$ & $7461$ & $7461$ \\
Number of EVs & $621$ & $621$ & $1552$ & $1552$ \\
Departure window & 0 h & 2 h & 0 h & 2 h \\ \hline
Total time to evacuate [s]  & $\phantom{0}19816$ & $\phantom{0}19920$ & & \\
Average speed [m/s] & $\phantom{000}5.32$ & $\phantom{00}10.58$ \\
Average duration [s] & $7712.58$ & $4851.47$ &  Same as Medium & Same as Medium  \\
Average waiting time [s] & $4789.48$ & $2397.17$ & 2040 All-at-once &  2040 2 hours  \\
Average time loss [s] & $6868.27$ & $4007.48$\\
Average departure delay [s] & $\phantom{0}543.12$ & $\phantom{000}2.79$ \\
\hline
\end{tabular}
\label{tab:small_scenario}
\vspace{-1em}
\end{table*}

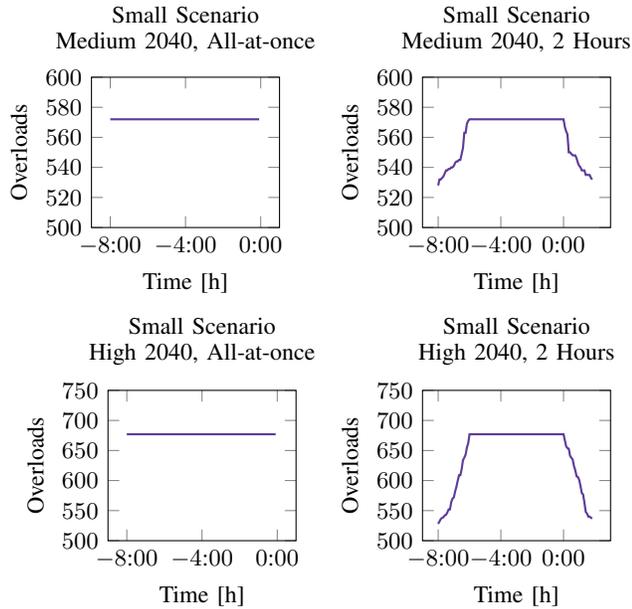
\begin{figure}
    \centering
%
%
%
\begin{tikzpicture}

\begin{axis}[%
width=2.5cm,
height=2cm,
scale only axis,
xmin=-9,
xmax=1,
xtick={0,-4,-8},
ymin=500,
ymax=600,
ytick={500,520,540,560,580,600},
ylabel={Overloads},
xlabel={Time [h]},
title style={align=center},
title={Small Scenario\\Medium 2040, All-at-once},
x filter/.code={\pgfmathparse{#1/60}},
xticklabel={ 
        \pgfmathsetmacro\hours{floor(\tick)}%
        \pgfmathsetmacro\minutes{(\tick-\hours)*0.6}%
        \pgfmathprintnumber{\hours}:\pgfmathprintnumber[fixed, fixed zerofill, skip 0.=true, dec sep={}]{\minutes}%
}
]
\addplot [mycolor1, thick, forget plot]
  table[row sep=crcr]{%
-480	572\\
-5	572\\
};
\end{axis}
\end{tikzpicture}
\begin{tikzpicture}
\begin{axis}[%
width=2.5cm,
height=2cm,
scale only axis,
xmin=-9,
xmax=3,
xtick={-8,-4,0},
xlabel={Time [h]},
ylabel={Overloads},
ymin=500,
ymax=600,
ytick={500,520,540,560,580,600},
title style={align=center},
title={Small Scenario\\Medium 2040, 2 Hours},
x filter/.code={\pgfmathparse{#1/60}},
xticklabel={ 
        \pgfmathsetmacro\hours{floor(\tick)}%
        \pgfmathsetmacro\minutes{(\tick-\hours)*0.6}%
        \pgfmathprintnumber{\hours}:\pgfmathprintnumber[fixed, fixed zerofill, skip 0.=true, dec sep={}]{\minutes}%
}
]
\addplot [mycolor1, thick, forget plot]
  table[row sep=crcr]{%
-480	528\\
-475	532\\
-470	532\\
-460	534\\
-450	538\\
-445	538\\
-440	539\\
-435	539\\
-430	540\\
-425	540\\
-420	541\\
-415	543\\
-410	544\\
-405	544\\
-400	545\\
-395	545\\
-385	553\\
-380	563\\
-375	563\\
-370	569\\
-365	571\\
-360	572\\
0	572\\
5	567\\
10	564\\
15	562\\
20	550\\
25	550\\
35	548\\
45	548\\
50	546\\
55	543\\
60	541\\
65	540\\
70	538\\
80	538\\
85	535\\
100	535\\
105	533\\
110	532\\
};
\end{axis}
\end{tikzpicture}

\vspace{0.1cm}

\begin{tikzpicture}
\begin{axis}[%
width=2.5cm,
height=2cm,
scale only axis,
xmin=-9,
xmax=1,
xtick={0,-4,-8},
xlabel={Time [h]},
ymin=500,
ymax=750,
ytick={500,550,600,650,700,750},
ylabel={Overloads},
title style={align=center},
title={Small Scenario\\High 2040, All-at-once},
x filter/.code={\pgfmathparse{#1/60}},
xticklabel={ 
        \pgfmathsetmacro\hours{floor(\tick)}%
        \pgfmathsetmacro\minutes{(\tick-\hours)*0.6}%
        \pgfmathprintnumber{\hours}:\pgfmathprintnumber[fixed, fixed zerofill, skip 0.=true, dec sep={}]{\minutes}%
}
]
\addplot [mycolor1, thick, forget plot]
  table[row sep=crcr]{%
-480	677\\
-5	677\\
};
\end{axis}
\end{tikzpicture}
\begin{tikzpicture}
\begin{axis}[%
width=2.5cm,
height=2cm,
scale only axis,
xmin=-9,
xmax=3,
xtick={0,-4,-8},
xlabel={Time [h]},
ylabel={Overloads},
ymin=500,
ymax=750,
ytick={500,550,600,650,700,750},
title style={align=center},
title={Small Scenario\\High 2040, 2 Hours},
x filter/.code={\pgfmathparse{#1/60}},
xticklabel={ 
        \pgfmathsetmacro\hours{floor(\tick)}%
        \pgfmathsetmacro\minutes{(\tick-\hours)*0.6}%
        \pgfmathprintnumber{\hours}:\pgfmathprintnumber[fixed, fixed zerofill, skip 0.=true, dec sep={}]{\minutes}%
}
]
\addplot [mycolor1, thick, forget plot]
  table[row sep=crcr]{%
-480	528\\
-470	536\\
-465	538\\
-460	539\\
-455	542\\
-450	543\\
-445	547\\
-440	552\\
-435	552\\
-430	564\\
-425	570\\
-420	572\\
-415	582\\
-410	590\\
-405	600\\
-400	609\\
-395	609\\
-390	619\\
-385	634\\
-380	638\\
-375	643\\
-370	655\\
-365	662\\
-360	677\\
0	677\\
10	657\\
15	654\\
20	653\\
25	641\\
35	635\\
40	624\\
45	617\\
50	607\\
55	606\\
60	601\\
65	591\\
70	579\\
75	577\\
80	563\\
85	548\\
90	545\\
95	540\\
100	540\\
105	538\\
110	537\\
};
\end{axis}
\end{tikzpicture}%
    \caption{Overloads per interval for medium and high EV penetration rates in the Small Scenario. In the left column, all the electric vehicles charge simultaneously. In the right column, the vehicle charging start times are randomly assigned over a two-hour window.}
    \label{fig:overloads_small_scenario}
    \vspace{-1em}
\end{figure}

\begin{figure}
    \centering
%
%
\definecolor{mycolor1}{rgb}{0.00000,0.44700,0.74100}%
\definecolor{mycolor2}{rgb}{0.85000,0.32500,0.09800}%
\definecolor{mycolor3}{rgb}{0.92900,0.69400,0.12500}%
\definecolor{mycolor4}{rgb}{0.49400,0.18400,0.55600}%
\begin{tikzpicture}

\begin{axis}[%
width=2.5cm,
height=2cm,
scale only axis,
xmin=-9,
xmax=1,
xtick={0, -4, -8},
ymin=100,
ymax=200,
ylabel={Overloads},
title style={align=center},
title={Medium 2040\\All-at-once},
enlarge x limits=false,
x filter/.code={\pgfmathparse{#1/60}},
xticklabel={ 
        \pgfmathsetmacro\hours{floor(\tick)}%
        \pgfmathsetmacro\minutes{(\tick-\hours)*0.6}%
        \pgfmathprintnumber{\hours}:\pgfmathprintnumber[fixed, fixed zerofill, skip 0.=true, dec sep={}]{\minutes}%
}
]
\addplot [mycolor1, thick, forget plot]
  table[row sep=crcr]{%
-480	133\\
-5	133\\
} ;
\addplot [mycolor2, thick, forget plot]
  table[row sep=crcr]{%
-480	142\\
-5	142\\
};
\addplot [mycolor3, thick, forget plot]
  table[row sep=crcr]{%
-480	182\\
-5	182\\
} ;
\addplot [mycolor4, thick, forget plot]
  table[row sep=crcr]{%
-480	115\\
-5	115\\
} ;
\end{axis}
\end{tikzpicture}
\begin{tikzpicture}
\begin{axis}[%
width=2.5cm,
height=2cm,
scale only axis,
xmin=-9,
xmax=3,
xtick={0, -4, -8},
ymin=100,
ymax=205,
ytick={100,120,140,160,180,200},
yticklabels={{},{},{},{},{},{}},
title style={align=center},
title={Medium 2040\\2 Hours},
x filter/.code={\pgfmathparse{#1/60}},
xticklabel={ 
        \pgfmathsetmacro\hours{floor(\tick)}%
        \pgfmathsetmacro\minutes{(\tick-\hours)*0.6}%
        \pgfmathprintnumber{\hours}:\pgfmathprintnumber[fixed, fixed zerofill, skip 0.=true, dec sep={}]{\minutes}%
}
]
\addplot [color=mycolor1, thick, forget plot]
  table[row sep=crcr]{%
-480	108\\
-475	112\\
-470	112\\
-460	114\\
-455	114\\
-450	113\\
-445	113\\
-440	114\\
-435	114\\
-430	113\\
-420	113\\
-415	115\\
-395	115\\
-390	118\\
-385	120\\
-380	128\\
-375	128\\
-370	133\\
-365	134\\
-360	133\\
0	133\\
5	131\\
10	128\\
15	129\\
20	117\\
25	117\\
30	116\\
35	117\\
45	117\\
50	118\\
55	116\\
70	113\\
80	113\\
85	112\\
100	112\\
105	113\\
110	112\\
};
\addplot [color=mycolor2, thick, forget plot]
  table[row sep=crcr]{%
-480	139\\
-460	139\\
-455	141\\
-450	144\\
-445	144\\
-440	142\\
-435	142\\
-430	144\\
-425	144\\
-420	145\\
-415	144\\
-410	145\\
-395	145\\
-390	143\\
-385	142\\
-380	143\\
-375	143\\
-370	141\\
-365	142\\
0	142\\
5	141\\
15	141\\
20	142\\
25	142\\
30	146\\
35	145\\
45	145\\
60	142\\
100	142\\
105	139\\
110	139\\
};
\addplot [color=mycolor3, thick, forget plot]
  table[row sep=crcr]{%
-480	172\\
-445	172\\
-440	174\\
-425	174\\
-420	172\\
-415	173\\
-405	173\\
-400	174\\
-395	174\\
-385	180\\
-380	181\\
-375	181\\
-370	183\\
-365	182\\
0	182\\
5	181\\
10	184\\
15	181\\
20	180\\
25	180\\
30	176\\
35	175\\
45	175\\
50	173\\
65	173\\
70	172\\
110	172\\
};
\addplot [color=mycolor4, thick,  forget plot]
  table[row sep=crcr]{%
-480	109\\
-425	109\\
-420	111\\
-375	111\\
-365	113\\
-360	115\\
0	115\\
5	114\\
10	111\\
80	111\\
85	109\\
110	109\\
};
\end{axis}
\end{tikzpicture}
\begin{tikzpicture}
\begin{axis}[%
width=2.5cm,
height=2cm,
scale only axis,
xmin=-9,
xmax=3,
xtick={0, -4, -8},
xlabel={Time [h]},
ymin=100,
ymax=205,
ytick={100,120,140,160,180,200},
ylabel={Overloads},
axis background/.style={fill=white},
title style={align=center},
title={High 2040\\All-at-once},
x filter/.code={\pgfmathparse{#1/60}},
xticklabel={ 
        \pgfmathsetmacro\hours{floor(\tick)}%
        \pgfmathsetmacro\minutes{(\tick-\hours)*0.6}%
        \pgfmathprintnumber{\hours}:\pgfmathprintnumber[fixed, fixed zerofill, skip 0.=true, dec sep={}]{\minutes}%
}
]
\addplot [color=mycolor1, thick, forget plot]
  table[row sep=crcr]{%
-480	184\\
-5	184\\
};
\addplot [color=mycolor2, thick, forget plot]
  table[row sep=crcr]{%
-480	167\\
-5	167\\
};
\addplot [color=mycolor3, thick, forget plot]
  table[row sep=crcr]{%
-480	201\\
-5	201\\
};
\addplot [color=mycolor4, thick, forget plot]
  table[row sep=crcr]{%
-480	125\\
-5	125\\
};
\end{axis}
\end{tikzpicture}
\begin{tikzpicture}
\begin{axis}[%
width=2.5cm,
height=2cm,
scale only axis,
xmin=-9,
xmax=3,
xtick={0, -4, -8},
xlabel={Time [h]},
ymin=100,
ymax=205,
ytick={100,120,140,160,180,200},
yticklabels={{},{},{},{},{},{}},
axis background/.style={fill=white},
title style={align=center},
title={High 2040\\ 2 Hours},
x filter/.code={\pgfmathparse{#1/60}},
xticklabel={ 
        \pgfmathsetmacro\hours{floor(\tick)}%
        \pgfmathsetmacro\minutes{(\tick-\hours)*0.6}%
        \pgfmathprintnumber{\hours}:\pgfmathprintnumber[fixed, fixed zerofill, skip 0.=true, dec sep={}]{\minutes}%
    }
]
\addplot [color=mycolor1, thick, forget plot]
  table[row sep=crcr]{%
-480	108\\
-475	112\\
-470	111\\
-465	113\\
-460	113\\
-450	115\\
-445	115\\
-440	119\\
-435	119\\
-430	126\\
-425	130\\
-420	131\\
-415	140\\
-410	142\\
-405	151\\
-400	157\\
-395	157\\
-390	162\\
-385	172\\
-380	170\\
-365	179\\
-360	184\\
0	184\\
5	179\\
10	176\\
15	178\\
20	178\\
25	170\\
35	168\\
40	158\\
45	153\\
50	149\\
55	152\\
65	146\\
70	138\\
75	136\\
80	127\\
85	115\\
90	114\\
100	114\\
105	112\\
110	113\\
};
\addplot [color=mycolor2, thick, forget plot]
  table[row sep=crcr]{%
-480	139\\
-475	139\\
-470	144\\
-465	142\\
-460	141\\
-455	142\\
-450	141\\
-445	145\\
-440	143\\
-435	143\\
-430	145\\
-425	146\\
-420	143\\
-415	142\\
-410	144\\
-405	139\\
-400	141\\
-395	141\\
-390	140\\
-385	143\\
-380	149\\
-375	148\\
-370	157\\
-365	158\\
-360	167\\
0	167\\
5	163\\
10	156\\
15	154\\
20	153\\
25	149\\
30	149\\
35	150\\
40	154\\
45	152\\
50	146\\
60	142\\
65	136\\
75	140\\
85	142\\
90	142\\
95	137\\
100	137\\
105	141\\
110	143\\
};
\addplot [color=mycolor3, thick, forget plot]
  table[row sep=crcr]{%
-480	172\\
-470	172\\
-460	176\\
-455	175\\
-450	176\\
-445	175\\
-440	178\\
-435	178\\
-425	180\\
-420	183\\
-415	184\\
-410	186\\
-405	192\\
-395	192\\
-390	195\\
-385	196\\
-380	196\\
-375	198\\
-370	198\\
-365	200\\
-360	201\\
0	201\\
5	200\\
10	200\\
15	197\\
20	200\\
25	201\\
30	199\\
35	196\\
40	192\\
45	192\\
50	193\\
55	191\\
60	193\\
65	193\\
70	188\\
75	186\\
80	181\\
85	177\\
90	175\\
95	176\\
105	176\\
110	172\\
};
\addplot [color=mycolor4, thick, forget plot]
  table[row sep=crcr]{%
-480	109\\
-460	109\\
-455	111\\
-450	111\\
-445	112\\
-435	112\\
-430	114\\
-425	114\\
-415	116\\
-410	118\\
-405	118\\
-400	119\\
-395	119\\
-390	122\\
-385	123\\
-380	123\\
-375	124\\
-370	124\\
-365	125\\
15	125\\
20	122\\
25	121\\
35	121\\
40	120\\
45	120\\
50	119\\
55	119\\
60	117\\
70	115\\
75	115\\
80	114\\
90	114\\
95	113\\
100	113\\
105	109\\
110	109\\
};
\end{axis}
\end{tikzpicture}

\begin{tikzpicture}
    \begin{customlegend}[legend entries={< $10$\%,$10$\% -- $50$\%,$50$\% -- $100$\%, >$100$\%}, legend columns=2]
    \addlegendimage{mycolor1, thick}
    \addlegendimage{mycolor2, thick}
    \addlegendimage{mycolor3, thick}
    \addlegendimage{mycolor4, thick}
    \end{customlegend}
\end{tikzpicture}\hspace{-1cm}
    \caption{Overloads under 10\% (blue), between 10\% and 50\% (orange), between 50\% and 100\% (yellow), and over 100\% (purple) for medium and high EV penetration rates in the small scenario. In the left column, all electric vehicles charge simultaneously. In the right column, the vehicle charging start times are randomly assigned over a two-hour window.}
    \label{fig:overloads_by_severity_small_scenario}
    \vspace{-1em}
\end{figure}
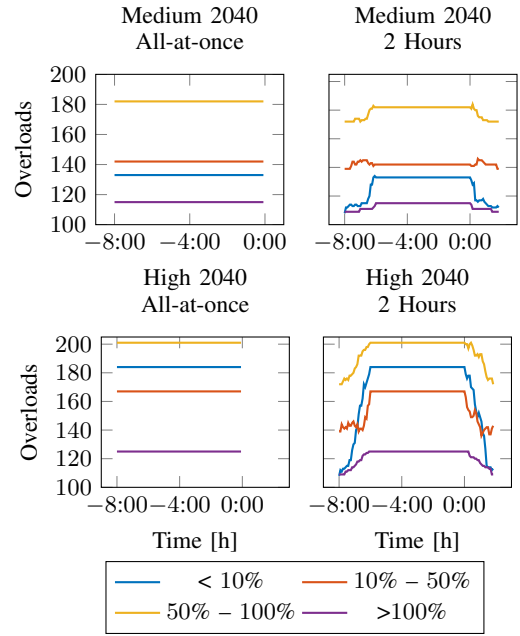

\begin{figure}
    \centering
    \includegraphics[width=0.48\textwidth]{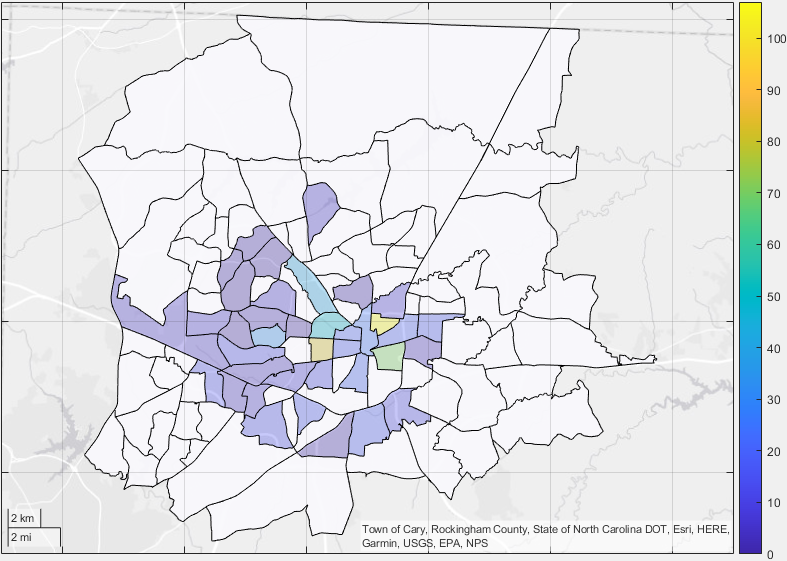}
    \caption{Map showing the distribution of the power grid overloads by TAZ during the last charging interval of the EVs in the scenario High 2040, All-at-once. White TAZs have no overloads in them, while the color gradient indicates the number of overloaded components in all other TAZs.}
    \label{fig:TAZ_overloads_map}
\end{figure}

To study the performance of the power grid, we will use the number of overloaded power grid components and the severity of these overloads to compare different evacuation charging schedules. A component is considered overloaded if the current flowing through it is above its specified normal current flow limit.  Figures~\ref{fig:overloads_small_scenario} and~\ref{fig:overloads_by_severity_small_scenario} demonstrate how the penetration rate of EVs and the timing of their charging can impact the number and severity of overloads, respectively. The left-hand figures depict the scenarios where all EVs charge simultaneously for eight hours, while the right-hand figures represent scenarios where the charging start times of all EVs are randomly assigned over a two-hour window. Figure~\ref{fig:overloads_small_scenario} shows that while simultaneous charging of all EVs reduces the overall charging time, this also results in more overloads over a longer period of time. In addition, the figures on the top and bottom illustrate that a higher number of EVs (High EV Penetration) can result in more overloads of power grid components than a moderate number of EVs (Medium EV Penetration). Furthermore, Figure~\ref{fig:overloads_by_severity_small_scenario} demonstrates how overloads of all severities increase the more EVs charge simultaneously. Overloads under 10\% see the highest increase, followed by overloads between 10\% and 50\%, overloads between 50\% and 100\%, and overloads over 100\%, respectively. 
These overloads are distributed across different levels and locations of the distribution network and the TAZs. Figure~\ref{fig:TAZ_overloads_map} shows a map of all TAZs and the number of power grid overloads occurring in each when all EVs are charging simultaneously in the High 2040 scenario. While many TAZs see a small number of overloads, most overloads are concentrated in the few TAZs which are evacuated in this scenario. To summarize, this small scenario shows that spreading out the departure of the vehicles reduces the duration of power grid overloads, while the total time to evacuate remains nearly the same (0.52\% increase).

\subsection{Large Evacuation Scenario}

For the large evacuation scenario, we consider all the TAZs displayed in Figure~\ref{fig:transportation_network}. We assign a random departure time for each vehicle within an eight-hour window. A summary of the number of vehicles considered and the performance of the transportation network is given in Table~\ref{tab:full_scenario}. As can be seen in Figure~\ref{fig:transportation_network}, the power network only covers part of the area, which means that only half of the total number of EVs will be included in the power grid simulator. We emphasize, however, that all TAZs are included in the traffic simulator.

Similar to the small scenario, the cumulative departure and arrival rate is shown in Figure~\ref{fig:cumarrfull}. This large scenario demonstrates the need to develop optimized evacuation plans since a naive routing where each vehicle is taking its shortest path with respect to uncongested does not seem to be a particularly efficient strategy. As can be seen in Table~\ref{tab:full_scenario}, the vehicles spend a lot of time waiting in the traffic network, and it is reasonable to believe that a more efficient routing strategy could alleviate part of this waiting time by making some vehicles take longer but less congested routes. 

\begin{figure}
    \centering
    \input{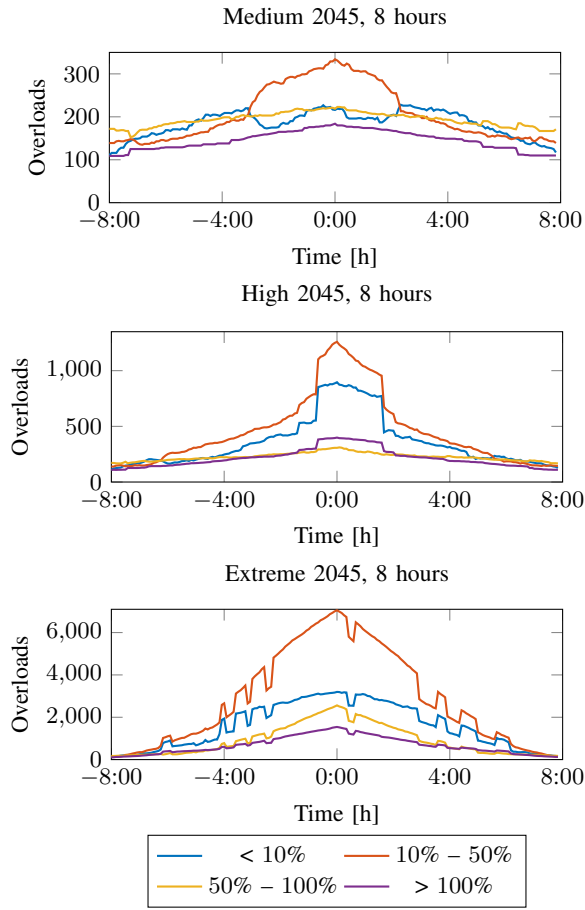}
    \caption{Overloads under 10\% (blue), between 10\% and 50\% (orange), between 50\% and 100\% (yellow), and over 100\% (purple) for medium, high, and extreme EV penetration rates in the large scenario. Vehicle charging start times are randomly assigned over an eight-hour window.}
    \label{fig:overloads_by_severity_large_scenario}
    \vspace{-1em}
\end{figure}

For the power grid simulations in the large scenario, we consider medium, high, and extreme EV penetration cases. Figure~\ref{fig:overloads_large_scenario} shows the number of overloaded components throughout the charging period. As expected, the number of overloaded components increases considerably as we increase the number of EVs in the system, with the extreme case having 1406\% and 401\% more overloads than the medium and high cases, respectively. The distribution of these overloads' severity is illustrated in Figure~\ref{fig:overloads_by_severity_large_scenario}. For all three penetration rates, overloads in the range of 10\% to 50\% account for the most of any category in this figure.

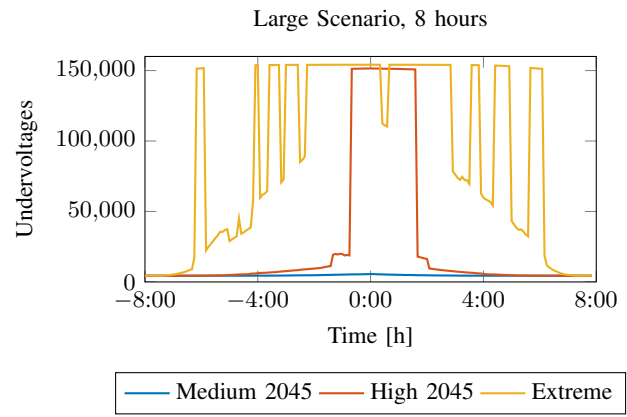
\begin{figure}
    \centering
%
%
\definecolor{mycolor1}{rgb}{0.00000,0.44700,0.74100}%
\definecolor{mycolor2}{rgb}{0.85000,0.32500,0.09800}%
\definecolor{mycolor3}{rgb}{0.92900,0.69400,0.12500}%
\begin{tikzpicture}

\begin{axis}[%
width=6cm,
height=3cm,
scale only axis,
xmin=-8,
xmax=8,
xtick={-8,-4,0,4,8},
xlabel={Time [h]},
ymin=0,
ymax=160000,
ylabel={Undervoltages},
title={Large Scenario, 8 hours},
legend style={at={(0.5,-0.4)},anchor=north},
legend columns=3,
yticklabel style={
        /pgf/number format/fixed,
        /pgf/number format/precision=5
},
scaled y ticks=false,
x filter/.code={\pgfmathparse{#1/60}},
xticklabel={ 
        \pgfmathsetmacro\hours{floor(\tick)}%
        \pgfmathsetmacro\minutes{(\tick-\hours)*0.6}%
        \pgfmathprintnumber{\hours}:\pgfmathprintnumber[fixed, fixed zerofill, skip 0.=true, dec sep={}]{\minutes}%
}
]
\addplot [color=mycolor1,thick]
  table[row sep=crcr]{%
-480	4419\\
-470	4419\\
-465	4420\\
-450	4420\\
-445	4421\\
-425	4421\\
-420	4423\\
-375	4423\\
-370	4424\\
-360	4424\\
-355	4426\\
-350	4426\\
-340	4428\\
-335	4428\\
-330	4429\\
-325	4429\\
-315	4431\\
-310	4431\\
-305	4432\\
-300	4438\\
-295	4438\\
-290	4439\\
-285	4442\\
-280	4444\\
-270	4444\\
-265	4451\\
-260	4455\\
-255	4460\\
-250	4460\\
-240	4464\\
-235	4470\\
-230	4473\\
-225	4473\\
-220	4484\\
-215	4483\\
-210	4493\\
-205	4493\\
-195	4529\\
-185	4543\\
-180	4556\\
-175	4595\\
-170	4607\\
-165	4648\\
-160	4671\\
-155	4683\\
-150	4711\\
-140	4729\\
-135	4743\\
-130	4749\\
-120	4773\\
-115	4788\\
-110	4831\\
-105	4886\\
-100	4893\\
-95	4952\\
-90	4980\\
-85	5014\\
-80	5101\\
-75	5112\\
-70	5128\\
-65	5161\\
-60	5176\\
-55	5220\\
-45	5339\\
-40	5346\\
-35	5381\\
-30	5387\\
-25	5398\\
-20	5442\\
-15	5458\\
-10	5505\\
-5	5527\\
0	5604\\
5	5576\\
10	5541\\
15	5453\\
20	5401\\
25	5341\\
30	5307\\
35	5231\\
40	5206\\
45	5206\\
50	5130\\
55	5100\\
60	5052\\
65	5052\\
70	4934\\
80	4924\\
90	4896\\
95	4820\\
100	4784\\
105	4772\\
110	4755\\
115	4750\\
120	4698\\
125	4692\\
130	4682\\
135	4654\\
140	4632\\
145	4632\\
150	4599\\
155	4587\\
160	4572\\
165	4566\\
170	4542\\
175	4538\\
180	4519\\
185	4509\\
190	4488\\
195	4474\\
200	4472\\
205	4471\\
210	4468\\
215	4466\\
225	4458\\
230	4452\\
235	4451\\
240	4449\\
245	4443\\
255	4441\\
260	4439\\
270	4439\\
275	4438\\
285	4434\\
290	4433\\
300	4433\\
310	4431\\
315	4429\\
330	4429\\
340	4427\\
345	4424\\
350	4423\\
365	4423\\
370	4422\\
380	4422\\
385	4421\\
400	4421\\
405	4420\\
440	4420\\
445	4419\\
470	4419\\
};
\addlegendentry{Medium 2045}
\addplot [color=mycolor2, thick]
  table[row sep=crcr]{%
-480	4419\\
-470	4419\\
-455	4422\\
-450	4422\\
-445	4423\\
-435	4423\\
-430	4424\\
-425	4424\\
-420	4427\\
-415	4429\\
-410	4430\\
-405	4430\\
-400	4431\\
-395	4434\\
-390	4435\\
-385	4437\\
-380	4441\\
-375	4442\\
-370	4450\\
-365	4455\\
-360	4473\\
-355	4481\\
-350	4494\\
-345	4508\\
-340	4539\\
-335	4555\\
-330	4606\\
-325	4647\\
-320	4678\\
-315	4714\\
-310	4763\\
-305	4839\\
-300	4962\\
-295	5023\\
-290	5127\\
-285	5200\\
-280	5285\\
-275	5328\\
-270	5382\\
-265	5419\\
-260	5557\\
-255	5851\\
-250	5920\\
-245	6002\\
-240	6175\\
-235	6365\\
-230	6459\\
-225	6492\\
-220	6618\\
-215	6707\\
-210	6843\\
-205	6923\\
-200	7030\\
-195	7252\\
-190	7396\\
-185	7505\\
-170	7986\\
-165	8141\\
-160	8274\\
-155	8376\\
-150	8596\\
-145	8830\\
-140	8927\\
-135	9038\\
-130	9220\\
-125	9356\\
-120	9503\\
-110	9835\\
-105	10016\\
-100	10406\\
-95	10735\\
-90	10992\\
-85	11281\\
-80	19289\\
-75	19856\\
-70	19470\\
-65	19999\\
-60	19739\\
-55	18785\\
-50	19196\\
-45	18848\\
-40	151231\\
-35	151286\\
-30	151312\\
-25	151345\\
-20	151363\\
-15	151423\\
-10	151472\\
-5	151496\\
0	151514\\
5	151458\\
10	151429\\
15	151413\\
20	151382\\
25	151349\\
30	151311\\
35	151266\\
40	151246\\
45	151232\\
50	151175\\
55	151160\\
60	151126\\
65	151102\\
70	151035\\
80	150943\\
85	150919\\
90	150846\\
95	150804\\
100	17940\\
105	17576\\
110	17081\\
115	16656\\
120	16264\\
125	9516\\
135	9073\\
145	8651\\
150	8419\\
155	8306\\
160	8090\\
165	7933\\
170	7802\\
175	7648\\
180	7419\\
190	7124\\
195	7015\\
200	6838\\
205	6705\\
210	6611\\
215	6528\\
220	6390\\
225	6171\\
230	6094\\
235	5974\\
240	5764\\
245	5510\\
250	5425\\
255	5368\\
260	5306\\
270	5210\\
275	5130\\
280	5090\\
285	4915\\
290	4866\\
295	4831\\
300	4761\\
305	4708\\
310	4626\\
315	4573\\
320	4563\\
325	4554\\
330	4524\\
335	4516\\
340	4500\\
345	4489\\
350	4479\\
355	4473\\
360	4466\\
365	4457\\
370	4453\\
375	4451\\
380	4444\\
385	4444\\
395	4434\\
400	4432\\
405	4429\\
410	4427\\
420	4427\\
425	4425\\
430	4424\\
435	4424\\
450	4421\\
455	4421\\
460	4419\\
470	4419\\
};
\addlegendentry{High 2045}
\addplot [color=mycolor3, thick]
  table[row sep=crcr]{%
-480	4419\\
-470	4421\\
-465	4423\\
-460	4424\\
-455	4434\\
-450	4439\\
-445	4456\\
-440	4534\\
-430	4767\\
-425	5043\\
-420	5208\\
-415	5512\\
-410	5922\\
-405	6243\\
-400	6627\\
-395	7159\\
-385	8403\\
-380	9036\\
-375	17104\\
-370	151317\\
-365	151460\\
-360	151580\\
-355	151711\\
-350	22453\\
-345	24920\\
-340	26900\\
-330	31223\\
-325	33113\\
-320	35633\\
-315	35495\\
-310	36873\\
-305	37409\\
-300	29029\\
-295	30140\\
-290	31323\\
-285	32374\\
-280	45090\\
-275	34190\\
-270	35168\\
-265	36352\\
-260	37600\\
-255	38631\\
-250	57708\\
-245	153760\\
-240	153798\\
-235	59733\\
-230	61956\\
-225	62731\\
-220	64508\\
-215	153906\\
-205	153930\\
-200	153937\\
-195	153938\\
-190	70700\\
-185	72892\\
-180	153985\\
-175	153994\\
-170	154005\\
-165	154024\\
-160	154032\\
-155	154034\\
-150	85068\\
-145	86637\\
-140	88914\\
-135	154051\\
-130	154053\\
-125	154056\\
-120	154061\\
-115	154061\\
-110	154078\\
-105	154078\\
-100	154086\\
-95	154093\\
-90	154094\\
-85	154094\\
-80	154096\\
-75	154097\\
-70	154099\\
-65	154113\\
-40	154113\\
-35	154118\\
20	154118\\
25	112491\\
30	111103\\
35	110253\\
40	154113\\
55	154113\\
60	154112\\
65	154110\\
75	154094\\
85	154094\\
90	154086\\
95	154080\\
105	154080\\
110	154066\\
125	154051\\
130	154044\\
135	154043\\
140	154043\\
150	154039\\
160	154023\\
165	154011\\
170	153997\\
175	78427\\
185	73875\\
190	72349\\
195	74583\\
200	72249\\
205	72223\\
210	69595\\
215	153883\\
220	153854\\
225	153834\\
230	153809\\
235	62605\\
240	59852\\
245	58578\\
250	57772\\
255	56738\\
260	54275\\
265	153532\\
270	153496\\
275	153429\\
280	153345\\
285	153287\\
290	153203\\
295	153081\\
300	43227\\
310	38463\\
315	36985\\
320	37274\\
325	35885\\
330	33558\\
335	32320\\
340	151871\\
345	151709\\
350	151606\\
355	151481\\
360	151344\\
365	151242\\
370	18559\\
375	11637\\
380	10291\\
385	9147\\
390	8148\\
395	7467\\
400	6731\\
405	6041\\
410	5562\\
415	5213\\
420	5001\\
425	4838\\
430	4624\\
435	4519\\
440	4477\\
445	4449\\
450	4437\\
455	4434\\
460	4424\\
465	4421\\
470	4420\\
};
\addlegendentry{Extreme}
\end{axis}

\end{tikzpicture}%
    \caption{Undervoltages per interval for Medium 2045, High 2045, and Extreme EV penetration rates in the large scenario. Vehicle charging start times are randomly assigned over an eight-hour window.}
    \label{fig:undervoltages_large_scenario}
    \vspace{-1em}
\end{figure}

In contrast to the small scenario, where charging a relatively small number of electric vehicles had almost no impact on the number of voltage violations, the large scenario showed a significant increase in undervoltage violations across all three cases. This is evident in Figure~\ref{fig:undervoltages_large_scenario}, which shows the number of undervoltage violations peaking at $5604$, $151514$, and $154118$  for the medium, high, and extreme cases, respectively. It is worth noting that Figures~\ref{fig:overloads_large_scenario}--\ref{fig:undervoltages_large_scenario} exhibit several discrete ``drops'' and ``jumps'' caused by the OpenDSS simulator's discrete controllers' efforts to minimize violations in the system. Figure~\ref{fig:controlVSnocontrol_large_scenario} depicts the reduction in undervoltages and overloaded components with and without these control efforts. These discrete changes in the system, especially transformer tap changes, help to reduce overloads and greatly decrease the number of voltage violations.

\begin{figure}
    \centering
\begin{tikzpicture}

\definecolor{darkgray176}{RGB}{176,176,176}
\definecolor{darkorange25512714}{RGB}{255,127,14}
\definecolor{steelblue31119180}{RGB}{31,119,180}
\begin{axis}[
tick align=outside,
tick pos=left,
x grid style={white!69.0196078431373!black},
xmin=0, xmax=50,
xtick style={color=black},
y grid style={white!69.0196078431373!black},
ymin=0, ymax=150000,
ytick style={color=black},
width=6cm, height=3.5cm,
xlabel={Time [h]},
ylabel={Vehicles},
title={Large Scenario, Medium 2045, 8 hours},
legend style={at={(0.5,-0.6)},anchor=north},
legend columns=2,
x filter/.code={\pgfmathparse{#1/3600}},
 xticklabel={ 
        \pgfmathsetmacro\hours{floor(\tick)}%
        \pgfmathsetmacro\minutes{(\tick-\hours)*0.6}%
        \pgfmathprintnumber{\hours}:\pgfmathprintnumber[fixed, fixed zerofill, skip 0.=true, dec sep={}]{\minutes}%
    },
yticklabel style={
        /pgf/number format/fixed,
        /pgf/number format/precision=5
},
scaled y ticks=false
]
\addplot [semithick, mycolor4]
table {%
0 2
176 841
2623 12826
3399 16543
3612 17624
4125 20144
4614 22533
4932 24038
5736 27883
5987 29115
6265 30432
6750 32896
7231 35241
7668 37316
10587 51372
14206 68242
14591 70020
22682 106874
22825 107495
23811 111832
24062 112970
24644 115451
24977 116895
25585 119498
26259 122518
26609 124047
27656 128529
28471 132073
28593 132593
28837 133526
32136 133707
32947 133837
35835 134175
37608 134514
38510 134710
40222 135158
42519 135483
43734 135581
45971 135755
47163 135872
47994 135921
49903 135986
78449 136606
79236 136635
80241 136684
82884 136793
84487 136842
85202 136878
86689 136956
87162 136963
89507 137012
92020 137063
94569 137112
95552 137141
98524 137249
100292 137369
100476 137399
104080 137592
104552 137619
109331 137859
110864 137932
111748 137981
112000 138008
139475 138508
145657 138557
146365 138564
149986 138613
151824 138637
160423 138686
163348 138695
172799 138695
};
\addlegendentry{Arrivals}
\addplot [semithick, mycolor1]
table {%
0 0
629 49
925 334
3648 6940
5236 10797
6411 13622
9123 20120
9789 21721
11427 25641
12238 27576
14385 32706
33890 79391
35201 81700
35578 82082
36035 82561
38230 84626
39542 85765
40841 86861
41556 87488
42219 88047
44326 89757
45705 90672
45958 90832
46440 91049
46807 91144
47842 91405
48369 91543
48933 91683
49882 91928
50998 92225
51473 92352
53970 93013
55408 93381
56757 93734
57520 93924
58081 94063
59622 94461
60272 94630
64910 95835
65189 95911
65881 96088
66680 96301
68697 96796
69979 97133
72553 97814
74290 98254
78161 99252
81483 100114
83058 100522
86228 101349
86909 101524
88309 101881
90660 102507
92290 102931
92783 103057
94344 103461
95849 103849
99099 104678
99781 104867
100632 105073
108276 107050
118517 109716
121631 110529
122280 110698
123046 110870
123814 111061
124866 111338
126486 111759
128565 112314
129233 112503
130345 112780
130609 112841
137023 114527
141702 115743
142600 115972
142970 116079
145629 116748
147866 117349
149052 117641
151011 118161
151708 118346
152930 118651
153615 118832
154846 119165
155989 119471
156641 119641
157411 119851
157940 119990
161902 121012
162283 121105
166393 122167
167320 122405
167983 122563
169136 122871
170389 123185
172799 123186
};
\addlegendentry{Departures}
\end{axis}

\end{tikzpicture}
    \caption{Cumulative departure and arrival curves for the large scenario where the vehicles depart during an eight-hour window.}
    \label{fig:cumarrfull}
    \vspace{-1em}
\end{figure}
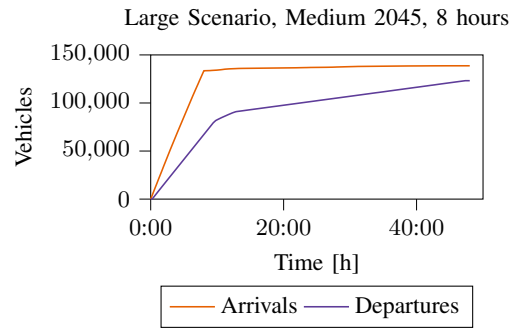

\begin{figure}
    \centering
%
%
\definecolor{mycolor1}{rgb}{0.00000,0.44700,0.74100}%
\definecolor{mycolor2}{rgb}{0.85000,0.32500,0.09800}%
\definecolor{mycolor3}{rgb}{0.92900,0.69400,0.12500}%
\begin{tikzpicture}

\begin{axis}[%
width=6cm, height=4cm,
xmin=-8,
xmax=8,
xlabel={Time [h]},
xtick={-8,-4,0,4,8},
ymin=0,
ymax=15000,
ylabel={Overloads},
title={Large Scenario, 8 hours},
legend style={at={(0.5,-0.5)},anchor=north},
legend columns=3,
x filter/.code={\pgfmathparse{#1/60}},
xticklabel={ 
        \pgfmathsetmacro\hours{floor(\tick)}%
        \pgfmathsetmacro\minutes{(\tick-\hours)*0.6}%
        \pgfmathprintnumber{\hours}:\pgfmathprintnumber[fixed, fixed zerofill, skip 0.=true, dec sep={}]{\minutes}%
},
yticklabel style={
        /pgf/number format/fixed,
        /pgf/number format/precision=5
},
scaled y ticks=false
]
\addplot [color=mycolor1, thick]
  table[row sep=crcr]{%
-480	531\\
-475	536\\
-465	536\\
-460	545\\
-455	549\\
-450	549\\
-445	560\\
-440	566\\
-435	570\\
-425	576\\
-420	577\\
-415	580\\
-410	584\\
-405	586\\
-400	586\\
-395	592\\
-390	600\\
-385	601\\
-380	603\\
-375	608\\
-370	614\\
-365	619\\
-360	625\\
-355	626\\
-350	629\\
-345	634\\
-340	635\\
-335	637\\
-330	638\\
-325	642\\
-315	646\\
-310	657\\
-305	658\\
-300	669\\
-295	674\\
-290	678\\
-285	681\\
-280	688\\
-275	697\\
-270	697\\
-265	703\\
-260	707\\
-250	717\\
-245	718\\
-235	730\\
-230	734\\
-225	734\\
-220	740\\
-215	747\\
-210	752\\
-205	752\\
-200	757\\
-195	760\\
-190	773\\
-185	778\\
-180	780\\
-175	790\\
-170	792\\
-165	798\\
-160	801\\
-155	802\\
-150	805\\
-145	810\\
-140	813\\
-130	823\\
-125	826\\
-120	827\\
-115	836\\
-110	840\\
-105	849\\
-100	853\\
-95	862\\
-90	864\\
-85	874\\
-80	881\\
-75	885\\
-70	888\\
-65	890\\
-60	904\\
-55	913\\
-50	914\\
-45	921\\
-40	926\\
-35	930\\
-30	932\\
-25	937\\
-20	941\\
-15	944\\
-10	953\\
-5	949\\
0	953\\
5	953\\
10	947\\
15	942\\
20	932\\
25	916\\
30	909\\
35	899\\
40	894\\
45	894\\
50	893\\
55	889\\
60	886\\
65	886\\
75	878\\
80	875\\
85	869\\
90	864\\
95	849\\
100	849\\
110	839\\
115	837\\
120	832\\
125	831\\
130	827\\
135	824\\
140	809\\
145	807\\
150	800\\
155	798\\
160	797\\
165	792\\
170	785\\
175	782\\
180	780\\
185	773\\
190	772\\
195	766\\
200	762\\
205	761\\
210	755\\
215	754\\
220	752\\
225	746\\
230	745\\
235	740\\
240	730\\
245	725\\
250	723\\
255	718\\
260	716\\
265	706\\
275	700\\
280	695\\
285	683\\
290	682\\
295	677\\
300	677\\
305	667\\
315	659\\
320	652\\
325	650\\
330	642\\
335	641\\
340	632\\
350	626\\
355	621\\
360	619\\
365	618\\
370	609\\
375	602\\
380	600\\
385	599\\
390	597\\
395	590\\
400	587\\
405	580\\
410	577\\
415	576\\
420	572\\
425	570\\
430	565\\
435	562\\
440	560\\
445	557\\
455	549\\
465	545\\
470	537\\
}; 
\addlegendentry{Medium 2045}
\addplot [color=mycolor2, thick]
  table[row sep=crcr]{%
-480	531\\
-475	542\\
-470	549\\
-465	555\\
-460	559\\
-455	572\\
-440	599\\
-435	610\\
-430	617\\
-425	625\\
-420	631\\
-410	651\\
-405	667\\
-400	684\\
-395	691\\
-390	700\\
-385	706\\
-380	714\\
-375	721\\
-370	743\\
-365	751\\
-360	763\\
-355	773\\
-350	784\\
-345	792\\
-340	796\\
-335	806\\
-330	822\\
-325	828\\
-320	842\\
-315	850\\
-310	871\\
-300	889\\
-295	902\\
-290	912\\
-285	930\\
-280	937\\
-275	942\\
-270	959\\
-265	970\\
-260	975\\
-255	988\\
-250	1006\\
-245	1013\\
-240	1029\\
-235	1046\\
-230	1060\\
-225	1078\\
-215	1110\\
-210	1120\\
-205	1144\\
-200	1182\\
-195	1199\\
-185	1244\\
-180	1257\\
-175	1280\\
-170	1289\\
-165	1297\\
-160	1314\\
-155	1329\\
-150	1351\\
-145	1370\\
-140	1384\\
-135	1400\\
-130	1420\\
-125	1437\\
-120	1453\\
-115	1471\\
-110	1483\\
-105	1503\\
-100	1525\\
-95	1541\\
-90	1564\\
-85	1581\\
-80	1775\\
-75	1797\\
-70	1812\\
-65	1836\\
-60	1852\\
-55	1872\\
-50	1906\\
-45	1912\\
-40	2630\\
-35	2660\\
-30	2680\\
-25	2704\\
-15	2777\\
-10	2810\\
-5	2837\\
0	2861\\
5	2812\\
10	2787\\
15	2754\\
20	2733\\
25	2697\\
30	2671\\
35	2631\\
40	2598\\
45	2556\\
50	2530\\
55	2500\\
60	2481\\
65	2456\\
70	2426\\
75	2405\\
80	2389\\
85	2363\\
90	2344\\
95	2329\\
100	1644\\
105	1628\\
110	1616\\
115	1596\\
120	1582\\
125	1441\\
130	1426\\
135	1408\\
140	1393\\
145	1388\\
150	1372\\
155	1359\\
160	1327\\
165	1311\\
170	1302\\
175	1273\\
180	1258\\
185	1238\\
190	1228\\
195	1217\\
200	1200\\
205	1185\\
210	1169\\
215	1160\\
220	1142\\
225	1137\\
230	1119\\
235	1104\\
240	1085\\
245	1065\\
250	1058\\
255	1046\\
260	1028\\
265	1014\\
275	988\\
280	952\\
285	938\\
290	926\\
295	898\\
300	894\\
305	876\\
310	865\\
315	856\\
320	849\\
325	827\\
330	820\\
335	807\\
340	795\\
345	790\\
350	778\\
355	767\\
360	752\\
365	738\\
370	737\\
375	724\\
380	713\\
385	707\\
390	702\\
395	688\\
400	678\\
405	670\\
410	658\\
415	645\\
425	625\\
430	621\\
435	616\\
445	590\\
450	584\\
455	577\\
460	566\\
465	562\\
470	548\\
};
\addlegendentry{High 2045}
\addplot [color=mycolor3, thick]
  table[row sep=crcr]{%
-480	537\\
-475	559\\
-470	594\\
-465	621\\
-460	656\\
-455	696\\
-450	749\\
-445	799\\
-435	862\\
-430	902\\
-425	959\\
-420	1006\\
-415	1073\\
-410	1110\\
-405	1165\\
-400	1214\\
-395	1267\\
-390	1342\\
-385	1393\\
-380	1467\\
-375	1665\\
-370	2417\\
-365	2533\\
-360	2635\\
-355	2756\\
-350	2072\\
-345	2166\\
-325	2440\\
-320	2534\\
-315	2618\\
-310	2688\\
-305	2785\\
-300	2664\\
-295	2771\\
-290	2885\\
-285	2992\\
-280	3287\\
-275	3146\\
-270	3246\\
-265	3351\\
-260	3489\\
-255	3591\\
-250	4019\\
-245	5800\\
-240	5977\\
-235	4412\\
-230	4546\\
-220	4796\\
-215	6862\\
-205	7144\\
-200	7316\\
-195	7515\\
-190	5677\\
-185	5841\\
-180	8067\\
-175	8282\\
-170	8439\\
-165	8644\\
-155	8986\\
-150	7009\\
-140	7419\\
-135	9706\\
-130	9906\\
-120	10279\\
-115	10436\\
-110	10604\\
-105	10785\\
-100	10977\\
-95	11190\\
-90	11371\\
-80	11751\\
-75	11955\\
-70	12118\\
-65	12309\\
-60	12458\\
-55	12657\\
-40	13165\\
-35	13278\\
-30	13409\\
-20	13719\\
-15	13897\\
-10	14040\\
-5	14203\\
0	14353\\
5	14209\\
10	14048\\
15	13902\\
20	13729\\
25	11409\\
35	11084\\
40	13068\\
45	12907\\
50	12770\\
60	12420\\
65	12204\\
70	12034\\
75	11813\\
80	11614\\
85	11447\\
90	11261\\
95	11119\\
100	10941\\
105	10777\\
110	10598\\
115	10411\\
120	10247\\
125	10040\\
130	9862\\
140	9456\\
145	9300\\
150	9155\\
155	8969\\
160	8773\\
165	8624\\
170	8460\\
175	6273\\
180	6086\\
185	5941\\
190	5804\\
195	5683\\
200	5549\\
205	5423\\
210	5241\\
215	6885\\
220	6702\\
225	6528\\
230	6339\\
235	4502\\
240	4335\\
245	4225\\
250	4093\\
255	3971\\
260	3834\\
265	5119\\
270	4991\\
275	4832\\
280	4697\\
285	4527\\
290	4393\\
295	4239\\
300	2959\\
305	2862\\
310	2750\\
325	2502\\
330	2398\\
335	2315\\
340	3069\\
345	2946\\
350	2846\\
355	2763\\
360	2628\\
365	2507\\
370	1728\\
375	1521\\
380	1454\\
385	1404\\
390	1351\\
395	1313\\
400	1269\\
405	1213\\
410	1139\\
415	1074\\
420	1041\\
425	1005\\
430	939\\
435	895\\
440	834\\
445	793\\
450	744\\
460	657\\
465	621\\
470	587\\
}; \addlegendentry{Extreme}

\end{axis}
\end{tikzpicture}%
    \caption{Overloads per interval for Medium 2045, High 2025, and Extreme EV penetration rates in the large scenario. Vehicle charging start times are randomly assigned over an eight-hour window.}
    \label{fig:overloads_large_scenario}
    \vspace{-1em}
\end{figure}

\begin{figure}
    \centering
%
%
\definecolor{mycolor1}{rgb}{0.00000,0.44700,0.74100}%
\definecolor{mycolor2}{rgb}{0.85000,0.32500,0.09800}%
\begin{tikzpicture}

\begin{axis}[%
width=6cm,
height=3cm,
scale only axis,
xmin=-8,
xmax=8,
xtick={-8,-4,0,4,8},
xlabel={Time [h]},
ymin=0,
ymax=160000,
ylabel={Undervoltages},
title={Extreme 2045, 8 hours},
yticklabel style={
        /pgf/number format/fixed,
        /pgf/number format/precision=5
},
scaled y ticks=false,
x filter/.code={\pgfmathparse{#1/60}},
xticklabel={ 
        \pgfmathsetmacro\hours{floor(\tick)}%
        \pgfmathsetmacro\minutes{(\tick-\hours)*0.6}%
        \pgfmathprintnumber{\hours}:\pgfmathprintnumber[fixed, fixed zerofill, skip 0.=true, dec sep={}]{\minutes}%
}
]
\addplot [color=mycolor1, forget plot,thick]
  table[row sep=crcr]{%
-480	4419\\
-470	4421\\
-465	4423\\
-460	4424\\
-455	4434\\
-450	4439\\
-445	4456\\
-440	4534\\
-430	4767\\
-425	5043\\
-420	5208\\
-415	5512\\
-410	5922\\
-405	6243\\
-400	6627\\
-395	7159\\
-380	9036\\
-375	17104\\
-370	151317\\
-365	151460\\
-360	151580\\
-355	151711\\
-350	22453\\
-345	24920\\
-340	26900\\
-330	31223\\
-325	33113\\
-320	35633\\
-315	35495\\
-310	36873\\
-305	37409\\
-300	29029\\
-295	30140\\
-290	31323\\
-285	32374\\
-280	45090\\
-275	34190\\
-270	35168\\
-265	36352\\
-260	37600\\
-255	38631\\
-250	57708\\
-245	153760\\
-240	153798\\
-235	59733\\
-230	61956\\
-225	62731\\
-220	64508\\
-215	153906\\
-205	153930\\
-200	153937\\
-195	153938\\
-190	70700\\
-185	72892\\
-180	153985\\
-175	153994\\
-170	154005\\
-165	154024\\
-160	154032\\
-155	154034\\
-150	85068\\
-145	86637\\
-140	88914\\
-135	154051\\
-130	154053\\
-125	154056\\
-120	154061\\
-115	154061\\
-110	154078\\
-105	154078\\
-100	154086\\
-95	154093\\
-90	154094\\
-85	154094\\
-80	154096\\
-75	154097\\
-70	154099\\
-65	154113\\
-40	154113\\
-35	154118\\
20	154118\\
25	112491\\
30	111103\\
35	110253\\
40	154113\\
55	154113\\
60	154112\\
65	154110\\
75	154094\\
85	154094\\
90	154086\\
95	154080\\
105	154080\\
110	154066\\
125	154051\\
130	154044\\
135	154043\\
140	154043\\
150	154039\\
160	154023\\
165	154011\\
170	153997\\
175	78427\\
185	73875\\
190	72349\\
195	74583\\
200	72249\\
205	72223\\
210	69595\\
215	153883\\
220	153854\\
225	153834\\
230	153809\\
235	62605\\
240	59852\\
245	58578\\
250	57772\\
255	56738\\
260	54275\\
265	153532\\
270	153496\\
275	153429\\
280	153345\\
285	153287\\
290	153203\\
295	153081\\
300	43227\\
310	38463\\
315	36985\\
320	37274\\
325	35885\\
330	33558\\
335	32320\\
340	151871\\
345	151709\\
350	151606\\
355	151481\\
360	151344\\
365	151242\\
370	18559\\
375	11637\\
380	10291\\
385	9147\\
390	8148\\
395	7467\\
400	6731\\
405	6041\\
410	5562\\
415	5213\\
420	5001\\
425	4838\\
430	4624\\
435	4519\\
440	4477\\
445	4449\\
450	4437\\
455	4434\\
460	4424\\
465	4421\\
470	4420\\
};
\addplot [color=mycolor2, forget plot,thick]
  table[row sep=crcr]{%
-480	154241\\
470	154241\\
};
\end{axis}
\end{tikzpicture}\hspace*{1cm}

\begin{tikzpicture} 
\begin{axis}[%
width=6cm,
height=3cm,
scale only axis,
xmin=-8,
xmax=8,
xtick={-8,-4,0,4,8},
xlabel={Time [h]},
ymin=0,
ymax=15000,
ylabel={Overloads},
axis background/.style={fill=white},
title={Extreme 2045, 8 hours},
yticklabel style={
        /pgf/number format/fixed,
        /pgf/number format/precision=5
},
scaled y ticks=false,
x filter/.code={\pgfmathparse{#1/60}},
xticklabel={ 
        \pgfmathsetmacro\hours{floor(\tick)}%
        \pgfmathsetmacro\minutes{(\tick-\hours)*0.6}%
        \pgfmathprintnumber{\hours}:\pgfmathprintnumber[fixed, fixed zerofill, skip 0.=true, dec sep={}]{\minutes}%
},
scaled y ticks=false,
legend style={at={(0.5,-0.4)},anchor=north},
legend columns=2
]
\addplot [color=mycolor1]
  table[row sep=crcr]{%
-480	537\\
-475	559\\
-470	594\\
-465	621\\
-460	656\\
-455	696\\
-450	749\\
-445	799\\
-435	862\\
-430	902\\
-425	959\\
-420	1006\\
-415	1073\\
-410	1110\\
-405	1165\\
-400	1214\\
-395	1267\\
-390	1342\\
-385	1393\\
-380	1467\\
-375	1665\\
-370	2417\\
-365	2533\\
-360	2635\\
-355	2756\\
-350	2072\\
-345	2166\\
-325	2440\\
-320	2534\\
-315	2618\\
-310	2688\\
-305	2785\\
-300	2664\\
-295	2771\\
-290	2885\\
-285	2992\\
-280	3287\\
-275	3146\\
-270	3246\\
-265	3351\\
-260	3489\\
-255	3591\\
-250	4019\\
-245	5800\\
-240	5977\\
-235	4412\\
-230	4546\\
-220	4796\\
-215	6862\\
-205	7144\\
-200	7316\\
-195	7515\\
-190	5677\\
-185	5841\\
-180	8067\\
-175	8282\\
-170	8439\\
-165	8644\\
-155	8986\\
-150	7009\\
-140	7419\\
-135	9706\\
-130	9906\\
-120	10279\\
-115	10436\\
-110	10604\\
-105	10785\\
-100	10977\\
-95	11190\\
-90	11371\\
-80	11751\\
-75	11955\\
-70	12118\\
-65	12309\\
-60	12458\\
-55	12657\\
-40	13165\\
-35	13278\\
-30	13409\\
-20	13719\\
-15	13897\\
-10	14040\\
-5	14203\\
0	14353\\
5	14209\\
10	14048\\
15	13902\\
20	13729\\
25	11409\\
35	11084\\
40	13068\\
45	12907\\
50	12770\\
60	12420\\
65	12204\\
70	12034\\
75	11813\\
80	11614\\
85	11447\\
90	11261\\
95	11119\\
100	10941\\
105	10777\\
110	10598\\
115	10411\\
120	10247\\
125	10040\\
130	9862\\
140	9456\\
145	9300\\
150	9155\\
155	8969\\
160	8773\\
165	8624\\
170	8460\\
175	6273\\
180	6086\\
185	5941\\
190	5804\\
195	5683\\
200	5549\\
205	5423\\
210	5241\\
215	6885\\
220	6702\\
225	6528\\
230	6339\\
235	4502\\
240	4335\\
245	4225\\
250	4093\\
255	3971\\
260	3834\\
265	5119\\
270	4991\\
275	4832\\
280	4697\\
285	4527\\
290	4393\\
295	4239\\
300	2959\\
305	2862\\
310	2750\\
325	2502\\
330	2398\\
335	2315\\
340	3069\\
345	2946\\
350	2846\\
355	2763\\
360	2628\\
365	2507\\
370	1728\\
375	1521\\
380	1454\\
385	1404\\
390	1351\\
395	1313\\
400	1269\\
405	1213\\
410	1139\\
415	1074\\
420	1041\\
425	1005\\
430	939\\
435	895\\
440	834\\
445	793\\
450	744\\
460	657\\
465	621\\
470	587\\
};
\addlegendentry{With voltage control}
\addplot [color=mycolor2, thick]
  table[row sep=crcr]{%
-480	537\\
-475	559\\
-470	594\\
-465	621\\
-460	656\\
-455	696\\
-450	749\\
-445	799\\
-435	862\\
-430	902\\
-425	959\\
-420	1006\\
-415	1073\\
-410	1110\\
-405	1165\\
-400	1214\\
-395	1267\\
-390	1342\\
-385	1393\\
-380	2207\\
-370	2417\\
-365	2533\\
-360	2635\\
-355	2756\\
-350	2839\\
-345	2166\\
-340	3050\\
-335	3164\\
-330	3291\\
-325	3388\\
-320	3532\\
-315	3651\\
-305	3942\\
-300	2880\\
-290	3129\\
-285	3213\\
-280	3287\\
-275	3146\\
-270	3246\\
-265	5141\\
-260	5298\\
-255	5439\\
-235	6168\\
-230	6368\\
-225	6529\\
-215	6862\\
-205	7144\\
-200	7316\\
-190	7706\\
-185	7874\\
-180	8067\\
-175	8282\\
-170	8439\\
-165	8644\\
-150	9159\\
-145	9344\\
-140	9545\\
-135	9706\\
-130	9906\\
-120	10279\\
-115	10436\\
-110	10604\\
-105	10785\\
-100	10977\\
-95	11190\\
-90	11371\\
-80	11751\\
-75	11955\\
-70	12118\\
-65	12309\\
-60	12458\\
-55	12657\\
-40	13165\\
-35	13278\\
-30	13409\\
-20	13719\\
-15	13897\\
-10	14040\\
-5	14203\\
0	14353\\
5	14209\\
10	14048\\
15	13902\\
25	13560\\
30	13399\\
35	13246\\
40	13068\\
45	12907\\
50	12770\\
60	12420\\
65	12204\\
70	12034\\
75	11813\\
80	11614\\
85	11447\\
90	11261\\
95	11119\\
100	10941\\
105	10777\\
110	10598\\
115	10411\\
120	10247\\
125	10040\\
130	9862\\
140	9456\\
145	9300\\
150	9155\\
155	8969\\
160	8773\\
165	8624\\
180	8136\\
185	7956\\
190	7815\\
200	7453\\
210	7067\\
230	6339\\
240	5932\\
245	5746\\
250	4093\\
255	3971\\
260	3834\\
265	5119\\
270	4991\\
275	4832\\
280	4697\\
285	4527\\
290	4393\\
295	4239\\
300	4077\\
305	3933\\
310	3797\\
315	3684\\
320	3548\\
325	3424\\
330	3319\\
335	3199\\
340	3069\\
345	2946\\
350	2846\\
355	2763\\
360	2628\\
365	2507\\
370	2451\\
375	2317\\
380	1454\\
385	1404\\
390	1351\\
395	1313\\
400	1269\\
405	1213\\
410	1139\\
415	1074\\
420	1041\\
425	1005\\
430	939\\
435	895\\
440	834\\
445	793\\
450	744\\
460	657\\
465	621\\
470	587\\
};
\addlegendentry{Without voltage control}
\end{axis}

\end{tikzpicture}%
    \caption{Comparison of undervoltages (left) and overloads (right) with and without allowing voltage control efforts in OpenDSS. Results where tap changes, voltage regulation, and capacitor switchings are allowed are shown in blue. Results where these control efforts are turned off are shown in orange.}
    \label{fig:controlVSnocontrol_large_scenario}
    \vspace{-1em}
\end{figure}

\begin{table*}
\centering
\caption{Statics about the transportation network for the large scenario}
\begin{tabular}{ccc} 
& Medium 2045 8 hours & High 2045 8 hours \\ \hline
Number of vehicles & $141046$  & $141046$ \\
Number of EVs & $13201$ & $29743$ \\
of which connected to the power grid & $7403$ & $16712$  \\
Departure window & 8 h & 8h \\ \hline
Total time to evacuate  & $>48$ h &  \\
Average speed [m/s] & $7.72$ &  \\
Average duration [s] & $27169.80$ & Same as Medium \\
Average waiting time [s] & $27169.80$ & 2045 8 hours \\
Average time loss [s] & $26353.44$ & \\
Average departure delay [s] & $1922.22$ &  \\
\hline
\end{tabular}
\label{tab:full_scenario}
\vspace{-1em}
\end{table*}

\subsection{Other Potential Applications}
While the current setup of the testbed is configured for evacuation scenarios, the testbed can easily be adapted to normal operational scenarios as well. For instance, one could compare weekday versus weekend transit behaviors by using parcel and zoning information to configure vehicles in the testbed to transit from homes to workplaces, schools, and commercial centers. The testbed could also be useful for studying the impacts of vehicle fleets (school buses, delivery vehicles, etc.) with different charging schedules.

To configure such scenarios, one would adjust the vehicle estimates for the parcel categories to make them match the scenario. Also, the flow and generation parts in the testbed workflow would need to be changed, as they are currently designed to route all the vehicles to one destination. For this, SUMO tools like activitygen\footnote{\url{https://sumo.dlr.de/docs/activitygen.html}} would be valuable.

\section{Conclusion and Future Work}\label{sec:conclusions}

In this work, we have presented a testbed that jointly simulates a transportation network and the power grid in the presence of electric vehicles that need to charge before departure. The testbed is constructed by linking together publicly available data from different sources, such as road map data from OpenStreetMap, parcel data from the State of North Carolina, and power grid data from the \mbox{SMART-DS} dataset. The open-source simulators SUMO and OpenDSS are then utilized by the testbed to simulate the transportation network and the power grid. To demonstrate how the testbed can be utilized, we performed simulations of both small- and large-scale evacuation scenarios. The simulations show the need to optimize evacuation scenarios while also demonstrating that synergies can be achieved by considering both systems jointly. Spreading out the vehicles' departure times may only cause small increases in the total evacuation time after the vehicles depart while substantially reducing the number and severity of overloaded power grid components.

Using the testbed presented in this paper, our ongoing work is continuing to analyze and optimize evacuation scenarios. Our ultimate goal is to optimally schedule vehicle charging as well as departure times and routes while jointly considering the capabilities of both the power grid and the transportation system. To this end, we envision extensions and enhancements of the testbed, such as modeling cascading failures in the electric grid model. Sufficiently large overloads of distribution lines and transformers may cause protection systems like fuses and breakers to operate, disconnecting the overloaded components, de-energizing portions of the grid, and potentially overloading other components. Modeling the behavior of protection systems would thus better represent possible outages during both normal and extreme operating conditions. Additionally, we aim to couple the vehicle energy consumption used for particular trips between SUMO and OpenDSS. For the evacuation scenarios that are the focus of the existing testbed, it suffices to fully charge the vehicles prior to departure. Conversely, trips during normal operating conditions may not require fully charging all vehicles prior to departure. Capturing this behavior would thus improve modeling accuracy during normal operating conditions.

\section*{Acknowledgements}
The authors would like to thank Yilun Chen for his support with the development and documentation of the testbed, Katelyn Iles for her help with obtaining vehicle estimates, Ravi Kodali for his support with power grid visualizations and running OpenDSS simulations, and Samuel Talkington for assisting with testing the tools in the testbed.

\bibliography{references}{}
\bibliographystyle{ieeetr}

\newpage

\vfill

\end{document}